\documentclass[iop]{emulateapj}
\usepackage{csvsimple}
\usepackage{adjustbox}
\usepackage{amsmath}
\usepackage{graphicx}
\usepackage[toc,page]{appendix}

\shorttitle{Molecular Emission in RY Lupi}
\shortauthors{Arulanantham et al.}

\begin{document}

\title{A UV-to-NIR Study of Molecular Gas in the Dust Cavity Around RY Lupi}

\author{N. Arulanantham and K. France}
\affil{Laboratory for Atmospheric and Space Physics, University of Colorado, 392 UCB, Boulder, CO 80303, USA}
\author{K. Hoadley}
\affil{Department of Astronomy, California Institute of Technology, 1200 East California Blvd., Pasadena, CA 91125, USA}
\author{C.F. Manara}
\affil{Science Support Office, Directorate of Science, European Space Research and Technology Centre (ESA/ESTEC), Keplerlaan 1, 2201 AZ, Noordwijk, The Netherlands}
\affil{European Southern Observatory, Karl-Schwarzschild-Str. 2, D-85748 Garching bei M\"{u}nchen, Germany}
\author{P.C. Schneider}
\affil{Hamburger Sternwarte, Gojenbergsweg 112, 21029 Hamburg, Germany}
\author{J.M. Alcal\'{a}}
\affil{INAF-Osservatorio Astronomico di Capodimonte, via Moiariello 16, 80131, Napoli, Italy}
\author{A. Banzatti}
\affil{Lunar and Planetary Laboratory, The University of Arizona, Tucson, AZ 85721, USA}
\author{H.M. G\"{u}nther}
\affil{MIT, Kavli Institute for Astrophysics and Space Research, 77 Massachusetts Avenue, Cambridge, MA 02139, USA}
\author{A. Miotello}
\affil{European Southern Observatory, Karl-Schwarzschild-Str. 2, D-85748 Garching bei M\"{u}nchen, Germany}
\affil{Leiden Observatory, Leiden University, Niels Bohrweg 2, 2333 CA Leiden, The Netherlands}
\author{N. van der Marel}
\affil{Herzberg Astronomy \& Astrophysics Programs, National Research Council of Canada, 5017 West Saanich Road, Victoria, BC, Canada V9E 2E7}
\author{E.F. van Dishoeck}
\affil{Leiden Observatory, Leiden University, P.O. Box 9513, NL-2300 RA Leiden, The Netherlands}
\affil{Max-Planck-Institut für Extraterrestrische Physik, Giessenbachstrasse 1, 85748 Garching, Germany}
\author{C. Walsh}
\affil{School of Physics and Astronomy, University of Leeds, Leeds LS2 9JT, UK}
\author{J.P. Williams}
\affil{Institute for Astronomy, University of Hawai'i at M¯anoa, Honolulu, HI, USA}

\begin{abstract}

We present a study of molecular gas in the inner disk $\left(r < 20 \, \text{AU} \right)$ around RY Lupi, with spectra from \emph{HST}-COS, \emph{HST}-STIS, and VLT-CRIRES. We model the radial distribution of flux from hot gas in a surface layer between $r = 0.1-10$ AU, as traced by Ly$\alpha$-pumped H$_2$. The result shows H$_2$ emission originating in a ring centered at $\sim$3 AU that declines within $r < 0.1$ AU, which is consistent with the behavior of disks with dust cavities. An analysis of the H$_2$ line shapes shows that a two-component Gaussian profile $\left(\text{FWHM}_{broad, H_2} = 105 \pm 15 \, \text{km s}^{-1}; \, \text{FWHM}_{narrow, H_2} = 43 \pm 13 \, \text{km s}^{-1} \right)$ is statistically preferred to a single-component Gaussian. We interpret this as tentative evidence for gas emitting from radially separated disk regions $\left( \left \langle r_{broad, H_2} \right \rangle \sim 0.4 \pm 0.1 \, \text{AU}; \, \left \langle r_{narrow, H_2} \right \rangle \sim 3 \pm 2 \, \text{AU} \right)$. The 4.7 $\mu$m $^{12}$CO emission lines are also well fit by two-component profiles $\left( \left \langle r_{broad, CO} \right \rangle = 0.4 \pm 0.1 \, \text{AU}; \, \left \langle r_{narrow, CO} \right \rangle = 15 \pm 2 \, \text{AU} \right)$. We combine these results with 10 $\mu$m observations to form a picture of gapped structure within the mm-imaged dust cavity, providing the first such overview of the inner regions of a young disk. The \emph{HST} SED of RY Lupi is available online for use in modeling efforts.

\end{abstract}


\section{Introduction}

The building blocks for planet formation are found in reservoirs of gas and dust around young stars. High resolution images acquired with the Atacama Large Millimeter Array (ALMA) have revealed the spatial extent of these protoplanetary disks (both radially and vertically) with high precision, showing prominent gaps in the dust continuum emission that potentially indicate clearing by young protoplanets (see e.g. \citet{HLTau2015, Isella2016, Andrews2016}). Gas appears to disperse quickly as the host star evolves \citep{Williams2011}, leaving less than 10 Myr \citep{Haisch2001} for young objects to form planetary cores and roughly establish the initial architecture of the system. It is therefore critical to study the mechanisms that deplete gas and dust from young disks in order to understand the resulting distribution of planets in extra-solar systems. 

Gas and small grains can be removed from protoplanetary disks through photoevaporation by far-ultraviolet (FUV), extreme ultraviolet (EUV), and X-ray radiation, stellar and disk winds, accretion onto the central star, outflows, and, in denser clusters, irradiation from external sources \citep{GH2009_photo, GH2009_phototime, Armitage2011, Alexander2014, Hartmann2016, Ercolano2017}. Despite the wealth of observational signatures from various atomic and molecular constituents of the gas disk, it is essential to study the physical properties of the most abundant component, hydrogen, in order to understand the behavior of the gas reservoir as a whole. Molecular hydrogen (H$_2$) lacks a permanent dipole moment, so pure rotational transitions are dipole-forbidden. H$_2$ can undergo quadrupole rotational transitions, but the large spacing between even the lowest energy levels makes it difficult to excite the molecules via collisions in cold midplane gas. 

As a heteronuclear molecule and the second most abundant molecular gas component, CO is typically used as a proxy \citep{Ansdell2016, Miotello2016}, but estimates of the H$_2$ abundance from these measurements rely on a H$_2$/CO ratio (see e.g. \citet{France2014}) that does not necessarily account for all the gas present in the system nor accurately treat freeze-out mechanisms in the disk midplane \citep{Long2017}. HD emission at 112 $\mu$m has also been used to estimate the total mass of the gas disk, since, as an isotopologue of H$_2$, it is expected to trace the distribution of hydrogen more closely than CO \citep{Bergin2013, McClure2016}. Alternatively, the population of H$_2$ in a hot $\left(T \sim 2000 \, \text{K} \right)$, thin layer at the surface of the disk can be observed directly through UV electronic transitions \citep{Herczeg2002, Herczeg2004, France2012_H2emission, Hoadley2015}. \citet{Herczeg2002} measured 146 UV-H$_2$ emission lines from TW Hya with \emph{HST}-STIS and found that the features are coincident with the star in velocity space, rather than spatially extended beyond the 0.05" resolution of the instrument, as would be expected for emission from an outflow. These observations indicate that the emitting H$_2$ is located in the inner regions of the protoplanetary disk (within $r \sim1.4$ AU at the distance of 56 pc to TW Hya), where gas temperatures can reach the 1500 K threshold required for Ly$\alpha$ fluorescence to take place \citep{Adamkovics2014, Adamkovics2016}.    

In addition to probing the hot H$_2$ in the inner disk, UV observations can be used to measure the properties of cooler molecular layers $\left(T \sim 300-600 \, \text{K} \right)$ of the disk via CO emission and absorption from the Fourth Positive band system $\left(A^1 \Pi-X^1\Sigma^+ \right)$ \citep{France2011_FUVcontII, Schindhelm2012_CO}. The CO emission lines are produced by the same mechanism as the UV-H$_2$ features, with C IV and Ly$\alpha$ emission pumping the gas to excited states \citep{France2011_FUVcontII}. Meanwhile, warm CO gas $\left( T \sim 300-1500 \text{ K} \right)$ is more favorably observed through the well-separated and strong rovibrational emission lines in the fundamental band at 4.7-5 $\mu$m, which have been studied in a large sample of $\left( > 60 \right)$ protoplanetary disks and probe their inner regions at 0.01-20 AU (e.g. \citet{Salyk2009, Salyk2011, Brown2013, Banzatti2015}). IR absorption at the same wavelengths provides additional constraints on the properties of CO in the disk atmosphere. By considering these UV and IR emission and absorption features together, we can begin to understand the radial structure of several different temperature regions within the inner molecular gas disk.  

In order to provide a more complete census on warm and hot molecular gas in planet-forming regions within protoplanetary disks, we present UV and IR observations of H$_2$ and CO in the young T Tauri system RY Lupi. By combining these inner disk gas tracers for the first time, we can map the radial structure in a region of the system where protoplanets may already be forming. We describe the target and observations in Section 2 and present our results from both wavelength regimes, including our modeling approach, in Section 3. Our results are evaluated in Section 4, where we consider the RY Lupi data in the context of larger surveys of the inner and outer regions of protoplanetary disks, including recent ALMA studies by \citet{Ansdell2016} and \citet{vanderMarel2018}. In a future work, this panchromatic approach will be extended to three other objects in the Lupus complex with different morphologies of gas and dust, allowing us to place RY Lupi on a spectrum of disk evolution that is derived from a co-evolving sample of systems.

\section{Target \& Observations}

\subsection{RY Lupi: An Unusual Object in the Lupus Complex}

RY Lupi is a particularly unusual member of the young (1-3 Myr), nearby \citep[$d \sim 151$ pc;][]{Gaia2016} Lupus cloud complex. Its near-UV $\left( \sim 3300 \text{ \AA} \right)$ continuum excess was used to determine an accretion luminosity of $\log{L_{acc} / L_{\odot}} = -0.9 \pm 0.25$ and mass accretion rate of $\log{\dot{M}_{acc}} = -8.2 \, M_{\odot}$ yr$^{-1}$ \citep{Alcala2017}, which makes it similar to a class II source with a full primordial disk. It was recently observed as part of a large ALMA survey that mapped the 890 $\mu$m dust continuum emission and the (3-2) features from $^{13}$CO and C$^{18}$O for 89 objects in the Lupus star-forming region \citep{Ansdell2016}. Although an infrared SED was previously used to classify RY Lupi as a primordial protoplanetary disk host \citep{KS2006}, the ALMA data show a distinct dust cavity in the mm-continuum emission with an outer radius of $\sim$50 AU \citep{Ansdell2016, vanderMarel2018}. \citet{vanderMarel2018} attribute the apparent discrepancy between the mid-IR and mm-wave observations to a misalignment between the inner and outer disk, requiring the inner disk to be close to face-on in order to reproduce the observed IR excess. This contrasts with the inclination of the outer disk, which has been constrained at a higher value of $68^{\circ}$ \citep{vanderMarel2018}.

The picture of a disk with components that are offset in inclination is supported by the system's unusual behavior in optical photometry, which shows variability over a period of $\sim$3.75 days that is accompanied by an increase in polarization when the star is faint \citep{Manset2009}. The observations have previously been explained as occultations by a warp in the inner disk that is co-rotating with the star \citep{Manset2009}, a model that was also invoked to describe the variability seen in AA Tau. However, this geometry is possible because of the nearly edge-on 85.6$^{\circ}$ inclination of the inner disk in RY Lupi, which was derived from Milgrom polarization models \citep{Manset2009} and is decidedly different from the value of 38$^{\circ}$ that best fits the mid-IR SED \citep{vanderMarel2018}. If the disk really is separated into two misaligned components, RY Lupi may be similar to the ``dipper disks" \citep[see e.g.][]{Ansdell2016_dippers}, which undergo photometric variability because of their geometries. At the time of our observations, synthetic photometry performed on our \emph{HST}-STIS spectrum of RY Lupi provided magnitudes of $U = 14.4$, $B = 13.6$, and $V = 12.5$. These brightnesses place the system at an intermediate phase between its faintest and brightest states, assuming its behavior is still well-represented by the light curve in \citet{Manset2009}. In this work, we aim to provide important constraints on the complex inner disk morphology of this peculiar system by unifying multiple tracers of its warm and hot gas.

\subsection{Observations}

RY Lupi was observed on March 16th, 2016 with the Cosmic Origins Spectrograph (COS) and the Space Telescope Imaging Spectrograph (STIS) onboard the \emph{Hubble Space Telescope (HST)} as part of a mid-cycle General Observer program (PID 14469; PIs: C.F. Manara, P.C. Schneider). Data were collected with three different observing modes of \emph{HST}-COS \citep[G140L $\lambda$1280, $R \sim 1500$, $t = 40.8$ m; G130M $\lambda$1291, $R \sim 16000$, $t = 10.9$ m; G160M $\lambda$1577, $R \sim 16000$, $t = 10.8$ m;][]{Green2012} as well as two different observing modes of \emph{HST}-STIS \citep[G430L $\lambda$4300, $t = 1$ m; G230L $\lambda$2375, $t = 40$ m; $R \sim 1000$;][]{Woodgate1997_overview, Woodgate1997_firstresults} over a total of five orbits. For the \emph{HST}-COS data, a final spectrum was produced by co-adding the original data products from the calibration pipeline \citep{Danforth2010}. Due to the uncertainty in the continuum flux uncertainties generated by this pipeline for low S/N sources, we treated the errors separately before including them in our modeling efforts (see Appendix A for details). 

\begin{figure*}
\centering
\includegraphics[width=0.85 \textwidth]
{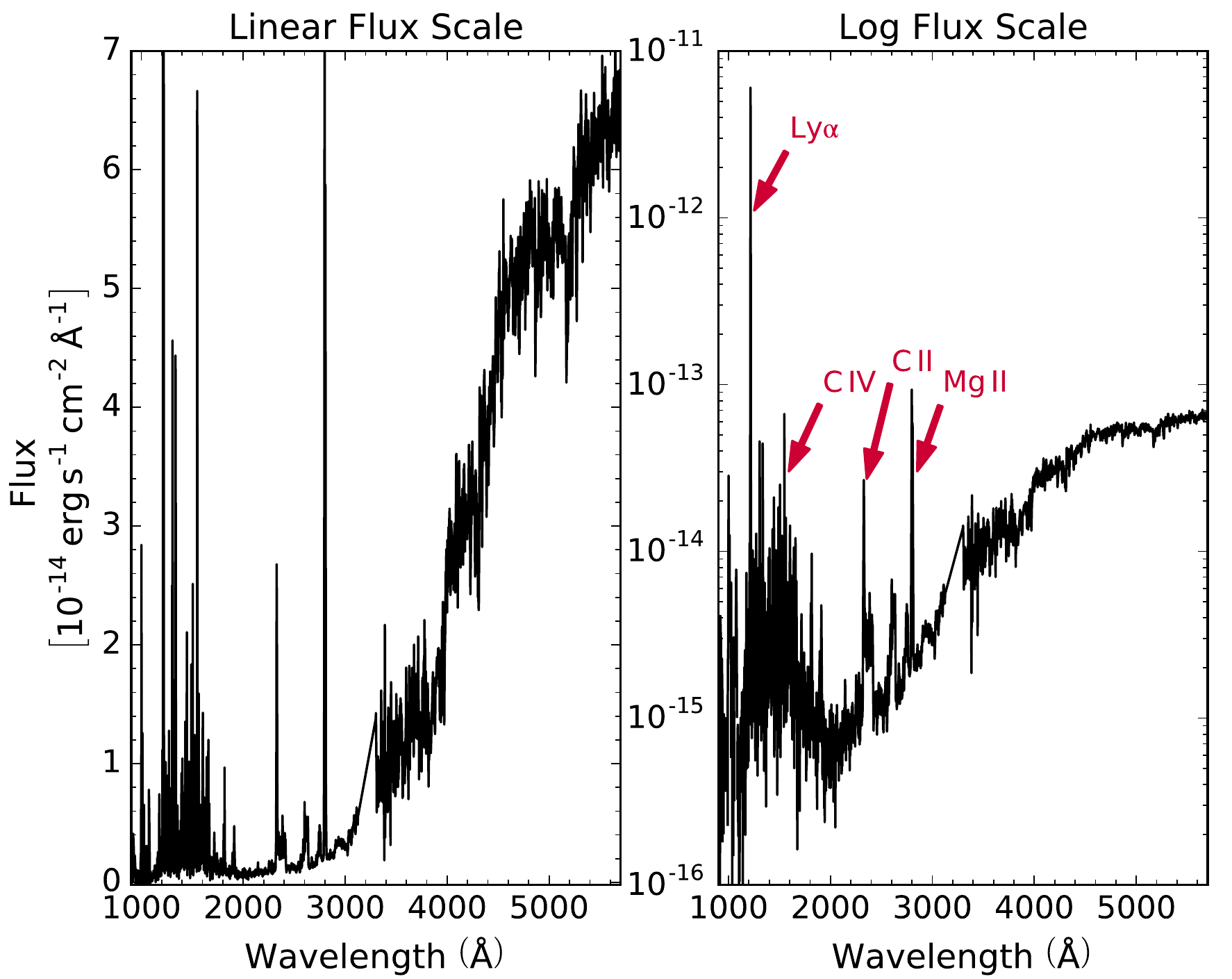}
\caption{A SED of RY Lupi, produced by stitching together spectra from five different observing modes of \emph{HST}-COS and \emph{HST}-STIS. Emission lines from Ly$\alpha$, C IV, C II, and Mg II are labeled \citep{Calvet2004}, and the 1600 \AA \, ``bump" is also prominent \citep{Bergin2004, Ingleby2009, France2011_FUVcontI, France2014, France2017}.}
\label{fullSED}
\end{figure*}

A full ultraviolet/optical SED was produced by stitching together the data from all five observing modes on \emph{HST}-COS and \emph{HST}-STIS that were used to observe RY Lupi (see Figure \ref{fullSED}), which have flux measurements that agree very well between modes with overlapping wavelengths. The data were rebinned to 5 \AA \, per pixel at wavelengths $\leq$1100 \AA \, to increase the signal-to-noise in the FUV, and the G130M and G160M spectra were smoothed with a 7 pixel boxcar kernel to make it easier to see the underlying continuum. For wavelength regions that were observed with multiple modes, the data from the higher resolution setting were used in the SED. The FUV continuum was then extracted by fitting a second-order polynomial to line-free regions of the spectrum between 1100-1715 \AA \, \citep{France2014}. A model Ly$\alpha$ profile was inserted in place of the observed line (see Section 3.1) in order to remove the effects of telluric emission and interstellar absorption. The full FUV radiation field is critical in dictating the overall disk chemistry, and our observed SED\footnote{Available at \url{http://cos.colorado.edu/\~kevinf/ctts\_fuvfield.html}} can be used in place of the simplified theoretical constraints typically used in modeling efforts.     

\section{Results}

\subsection{UV Observations of Ly$\alpha$-Pumped H$_2$ Emission Lines}

We detect fluorescent emission from H$_2$ molecules in a hot layer at the disk surface. Ly$\alpha$ photons pump this gas into excited rovibrational states, denoted $m$, within the $B^1 \Sigma_{u}^{+}$ electronic state \citep{Herczeg2002}. Each pumping wavelength along the Ly$\alpha$ profile produces a progression of emission lines, consisting of all transitions from $m$ $\left(\left[\nu', J' \right] \right)$ to rovibrationally excited $\left[\nu'', J'' \right]$ levels of the ground electronic state, $X^1 \Sigma_{g}^{+}$. Our analysis here is focused on H$_2$ features in the \emph{HST}-COS G160M data $\left(\Delta v \sim 17 \text{ km/s} \right)$, which extend to wavelengths that are minimally impacted by self-absorption \citep[$\lambda > 1450 \text{\, \AA;}$][]{McJunkin2016}. 

Fluxes from the H$_2$ profiles were first measured by fitting a Gaussian profile superimposed onto a linear continuum to each emission line\footnote{Fluxes were measured using a GUI (SELFiE: \textbf{S}TIS/COS \textbf{E}mission \textbf{L}ine \textbf{Fi}tting and \textbf{E}xtraction), which was developed for interactively fitting spectral lines in Python with the non-linear least squares algorithm \emph{scipy.optimize.curve\_fit}. The code is available at \url{https://github.com/narulanantham/SELFiE}}. The Gaussian first had to be convolved with a wavelength-dependent line-spread function (LSF) to account for the effects of wave-front errors induced by the primary and secondary mirrors on \emph{HST} \citep{France2012_H2emission}. This model was applied to individual emission lines from 12 progressions with pumping wavelengths along the Ly$\alpha$ profile. After de-reddening the spectrum using the optical extinction \citep[$A_V = 0.4;$][]{Alcala2017}, the flux $\left(F_{mn} \right)$ from each emission line in a given progression, scaled by its transition rate relative to all other pathways to the ground state $\left(B_{mn} \right)$, can be summed as 
\begin{equation}
F_m \left(H_2 \right) = \frac{1}{N} \sum_{n=1}^N \left( \frac{F_{mn}}{B_{mn}} \right)
\end{equation}
to yield the total flux from molecules in the number of states $\left(N \right)$ that were excited by a single Ly$\alpha$ pumping wavelength. Most of the lines in the [3,13], [4,13], [3,0], [2,15], and [0,3] progressions are indistinguishable from the continuum (see Table \ref{H2progfluxes}), so the corresponding estimates of $F_{mn}$ are upper limits. 

\begin{deluxetable*}{ccccc}
\tabletypesize{\small}
\tablewidth{0.7 \linewidth}
\tablecaption{Progression Fluxes for Ly$\alpha$ Pumped H$_2$ Emission \label{H2progfluxes}
}
\tablehead{
\colhead{Pumping Wavelength} & \colhead{Progression} & \colhead{$F_{m} \left(H_2 \right)$} & \colhead{$\left< \text{FWHM} \right>$} & \colhead{\tablenotemark{a}$\left< R_{H_2} \right>$}  \\ 
\colhead{\AA} & \colhead{$\left[ \nu', J' \right]$} & \colhead{$10^{-12}$ erg s$^{-1}$ cm$^{-2}$} & \colhead{km s$^{-1}$} & \colhead{AU} \\
}
\startdata
1213.36 & $\left[3,13 \right]$ & $\leq 0.6$ & - & - \\
1213.68 & $\left[4,13 \right]$ & $\leq 10$ & - & - \\
1214.47 & $\left[3,16 \right]$ & $1.4 \pm 0.2$ & $50 \pm 20$ & $3 \pm 2$ \\
1214.78 & $\left[4,4 \right]$ & $\leq 4.7$ & - & - \\
1215.73 & $\left[1,7 \right]$ & $2.23 \pm 0.07$ & $48 \pm 2$ & $2.6 \pm 0.7$ \\ 
1216.07 & $\left[1,4 \right]$ & $3.74 \pm 0.05$ & $51 \pm 1$ & $2.4 \pm 0.6$ \\
1217.04 & $\left[3,0 \right]$ & $\leq 16$ & - & - \\
1217.21 & $\left[0,1 \right]$ & $0.93 \pm 0.07$ & $46 \pm 9$ & $3 \pm 1$ \\
1217.64 & $\left[0,2 \right]$ & $0.80 \pm 0.05$ & $53 \pm 7$ & $2.2 \pm 0.8$ \\
1217.9 & $\left[2,12 \right]$ & $0.53 \pm 0.08$ & $50 \pm 17$ & $2 \pm 1$ \\
1218.52 & $\left[2,15 \right]$ & $\leq 1.3$ & - & - \\
1219.09 & $\left[0,3 \right]$ & $\leq 0.21$ & - & - \\
\enddata
\end{deluxetable*}

\begin{figure*}
\centering
\includegraphics[width=0.9 \textwidth]
{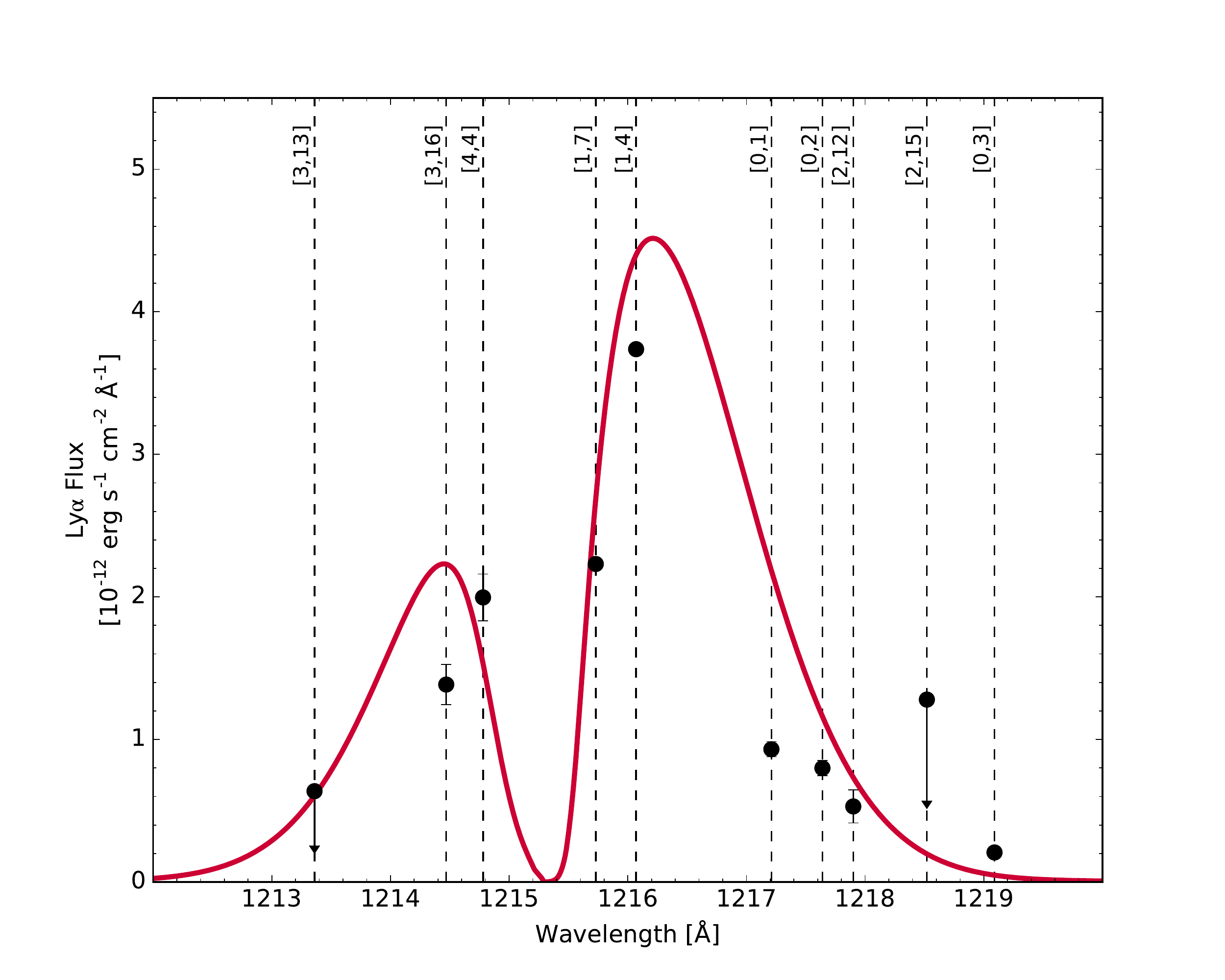}
\caption{H$_2$ fluxes (black circles) were used to estimate the outflow-absorbed (red) Ly$\alpha$ profile in RY Lupi, which represents the radiation field seen by the hot molecular layer at the surface of the disk, where the observed H$_2$ fluorescence originates. This estimated profile has a width of $\sim$600 km/s, an integrated line flux of $\sim$$10^{-11}$ erg s$^{-1}$ cm$^{-2}$, and an outflow velocity of -225 km/s. After removing the geocoronal emission between 1214.7 and 1216.7 \AA, the ratio of observed to reconstructed Ly$\alpha$ flux is 0.05.}
\label{LyAreconstfig}
\end{figure*}

In order to accurately model the physical properties of the emitting gas, we must include the Ly$\alpha$ profile, as seen at the disk surface, as the primary excitation source. However, the observed Ly$\alpha$ line is attenuated by interstellar H I and telluric emission and cannot be used directly. The intrinsic profile can be reconstructed from the measured H$_2$ emission fluxes, as carried out by \citet{Schindhelm2012} and \citet{France2014} for a sample of classical T Tauri stars. Their catalog of Ly$\alpha$ profiles were compared to our data, showing that the width of the observed Ly$\alpha$ line in RY Lupi appears to be most similar to the profiles from V4046 Sgr and RECX-11. A superposition of these two sources was chosen as the ``reconstructed" line and scaled to the distance of RY Lupi (see Figure \ref{LyAreconstfig}). 

To verify that this is a reasonable estimate of the intrinsic and outflow-absorbed Ly$\alpha$ flux from the system, the adopted profile was used to generate a 1-D model of the H$_2$ fluorescence spectrum \citep{McJunkin2016}. The model was able to reasonably reproduce the observed H$_2$ emission lines in the progressions given by \citet{France2012_H2emission}, within the temperature range expected for this hot gas $\left (T \sim 1500-2500 \, \text{K} \right)$. We find that the [0,1] and [3,16] progressions, which both have reasonably well-defined emission lines, only contain about half the total flux that is expected from the selected outflow-absorbed profile at these wavelengths. Furthermore, the [4,4] progression flux is $\sim$3 times higher than expected, a result also seen by \citet{McJunkin2016}. It returns to the expected level when the (4$-$9)P(5) 1526.55 \AA \, feature, which is barely distinguishable from the continuum, is dropped from the total flux calculation.

\begin{figure*}
\centering
\includegraphics[width=0.9\textwidth]
{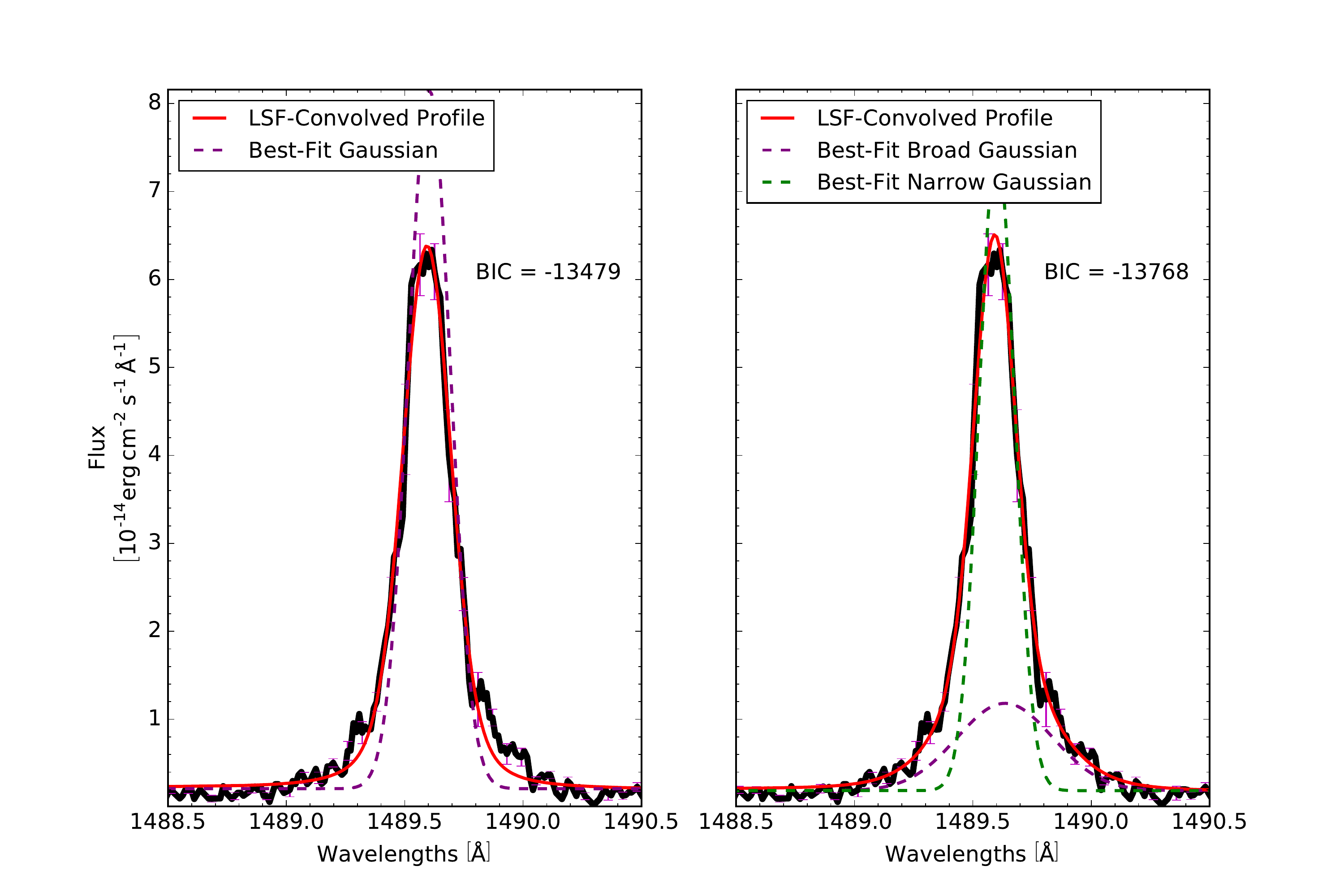}
\caption{A two-component Gaussian fit to the strongest fluorescent UV-H$_2$ emission features (right) gives a statistically better fit to the line profiles than a single-component Gaussian (left). This tentatively implies that the H$_2$ features may be a superposition of emission from radially separated regions of the disk, as observed in the IR-CO lines \citep[see e.g. Figure 6 and][]{Banzatti2015}. The dashed lines shown in each subplot are obtained from the best-fit Gaussian parameters, which are convolved with the instrument line-spread function (solid, blue) before fitting to the data in order to mimic the redistribution of flux from the peak to the wings. For the (1-7) R(3) line at 1489.57 \AA \, shown here, the difference in the Bayesian Information Criterion (BIC) that was calculated for each model is $\Delta BIC = 289 >> 10,$ where 10 is the expected threshold for a statistically significant difference between models \citep[see e.g.][]{RM2016, Manara2017}.}
\label{twocompgauss}
\end{figure*}

The H$_2$ profiles with the highest S/N show slightly more emission in the line wings than what is expected from a Gaussian profile produced by gas in Keplerian rotation (see Figures \ref{twocompgauss} and \ref{H2emissionfits}). Similar line morphologies have been observed in 4.7 $\mu$m CO ro-vibrational emission in the $v = 1-0$ band \citep{Bast2011, Brown2013}, showing in some cases a noticeable discontinuity in the line profile between the broad wings and a narrow central peak \citep{Banzatti2015}. These shapes could in some cases be produced by a line brightness profile that deviates from a Gaussian \citep[see e.g.][]{Bast2011}, with a possible contribution from a slow disk wind \citep{Pontoppidan2011}, and in other cases potentially indicate a depletion of CO gas in a gap at disk radii corresponding to the intermediate velocities between the broad and narrow components \citep{Banzatti2015}. This latter scenario is consistent with the strong discontinuity observed in the IR-CO line profiles of RY Lupi (see Section 3.3 and Figure \ref{H2COcomp}). Below, we assume that a radial gap, although much narrower, could also explain the UV-H$_2$ line profiles. 

A two-component model consisting of broad and narrow LSF-convolved Gaussians was fit to the four strongest lines in the [1,4] progression (see Figure \ref{twocompgauss}). The best-fit average FWHM of the broad component was $\text{FWHM}_{broad, H_2} = 105 \pm 15$ km s$^{-1}$, while the narrow component had a width of $\text{FWHM}_{narrow, H_2} = 43 \pm 13$ km s$^{-1}$. Given the stellar mass \citep[$1.4 \, M_{\odot}$;][]{Manset2009, Alcala2017} and inner disk inclination \citep[$85.6^{\circ}$;][]{Manset2009}, the FWHMs from the broad and narrow Gaussian components can be converted to average H$_2$ radii \citep{France2012_H2emission} 
\begin{equation}
\left< R_{H_2} \right> = G M_{\ast} \left( \frac{2 \sin i_{inner}}{\text{FWHM}} \right)^2 
\end{equation}
of $\left \langle r_{broad, H_2} \right \rangle = 0.4 \pm 0.1$ AU and $\left \langle r_{narrow, H_2} \right \rangle = 3 \pm 2$ AU. To determine whether the one- or two-component profile provides a better description of the data, the Bayesian Information Criterion (BIC) 
\begin{equation}
BIC = p_j \log n - 2 \mathcal{L} \left( \hat{\theta_j} \right)
\end{equation}
was computed for both models, where $p_j$ is the number of parameters in the model, $n$ is the number of data points, and $\hat{\theta_j}$ is the set of parameter values that maximize the likelihood function \citep{Schwarz1978}. The second model yielded a significantly lower BIC \citep[$\Delta BIC > 10$;][]{RM2016, Manara2017}, implying that the two-component fit is statistically preferred (see Table \ref{H2BIC}). However, we note that our estimate of the average emission radius for the narrow component has a large uncertainty. We discuss the difference between the two components in the context of other observational metrics of the disk in Section 4.2 but caution the reader against interpreting $\left \langle r_{broad, H_2} \right \rangle$ and $\left \langle r_{narrow, H_2} \right \rangle$ in an absolute sense.

\begin{deluxetable}{cccc}
\tablecaption{Bayesian Information Criterion (BIC) for H$_2$ Emission Line Fits \label{H2BIC}
}
\tablewidth{0.9 \linewidth}
\tabletypesize{\scriptsize}
\tablehead{
\colhead{Wavelength} & \colhead{\tablenotemark{a}$BIC_1$} & \colhead{\tablenotemark{b}$BIC_2$} & $\Delta BIC$ \\
\colhead{\AA} & & & 
}
\startdata
1431.01 & -19567 & -19584 & 18 \\
1446.12 & -17789 & -17962 & 173 \\
1489.57 & -13479 & -13768 & 289 \\
1504.76 & -13836 & -14228 & 391 \\
\enddata
\tablenotetext{a}{BIC calculated for the single component line profile}
\tablenotetext{b}{BIC calculated for the line profile with a broad and a narrow component}
\end{deluxetable}

\subsection{Radial Distribution of Ly$\alpha$-Pumped UV-H$_2$ Emission}

In order to map the spatial location of the hot, fluorescent UV-H$_2$, we applied the 2-D radiative transfer modeling approach from \citet{Hoadley2015} to emission lines in the [1,4], [1,7], and [0,2] progressions, which have the strongest S/N. The gas is assumed to be a thin, inclined $\left(i_{inner} = 85.6^{\circ} \right)$ surface layer in Keplerian rotation 
\begin{equation}
v_{\phi} \left( r \right) = \sqrt{ \frac{G M_{\ast}}{r}},
\end{equation}
where the bulk motion of the gas dominates the line widths. We also assume local thermodynamic equilibrium (LTE) conditions for the ground states. The temperature distribution is described by a power law of index $q$, normalized to $T_{1 AU}$ at a distance of 1 AU from the central star, 
\begin{equation}
T \left( r \right) = T_{1 \, \text{AU}} \left( \frac{r}{1 \, \text{AU}} \right)^{-q}
\end{equation}
and is taken as azimuthally symmetric and isothermal with height in the atmosphere. A power-law of index $\gamma$ is combined with an exponential cutoff beyond some characteristic radius $r_c$ to describe the surface density distribution, 
\begin{equation}
\Sigma \left(r \right) = \Sigma_c \left( \frac{r}{r_c} \right)^{-\gamma} \exp \left[- \left( \frac{r}{r_c} \right)^{2- \gamma} \right],
\end{equation}
which is integrated over radius to get a total mass of hot H$_2$ $\left(M_{H_2, hot}\right)$. The final emission line profiles are a collapsed view of the entire disk, which are produced from the model by summing the flux at each $\left(r, z \right)$ corresponding to a given velocity. Each profile is convolved with the \emph{HST}-COS LSF corresponding to the central wavelength of the line before comparison with the observed data. More detailed descriptions of the modeling approach are provided in Appendix B of this work and \citet{Hoadley2015}. 

\begin{figure*}
\centering
\includegraphics[width=0.9\textwidth]
{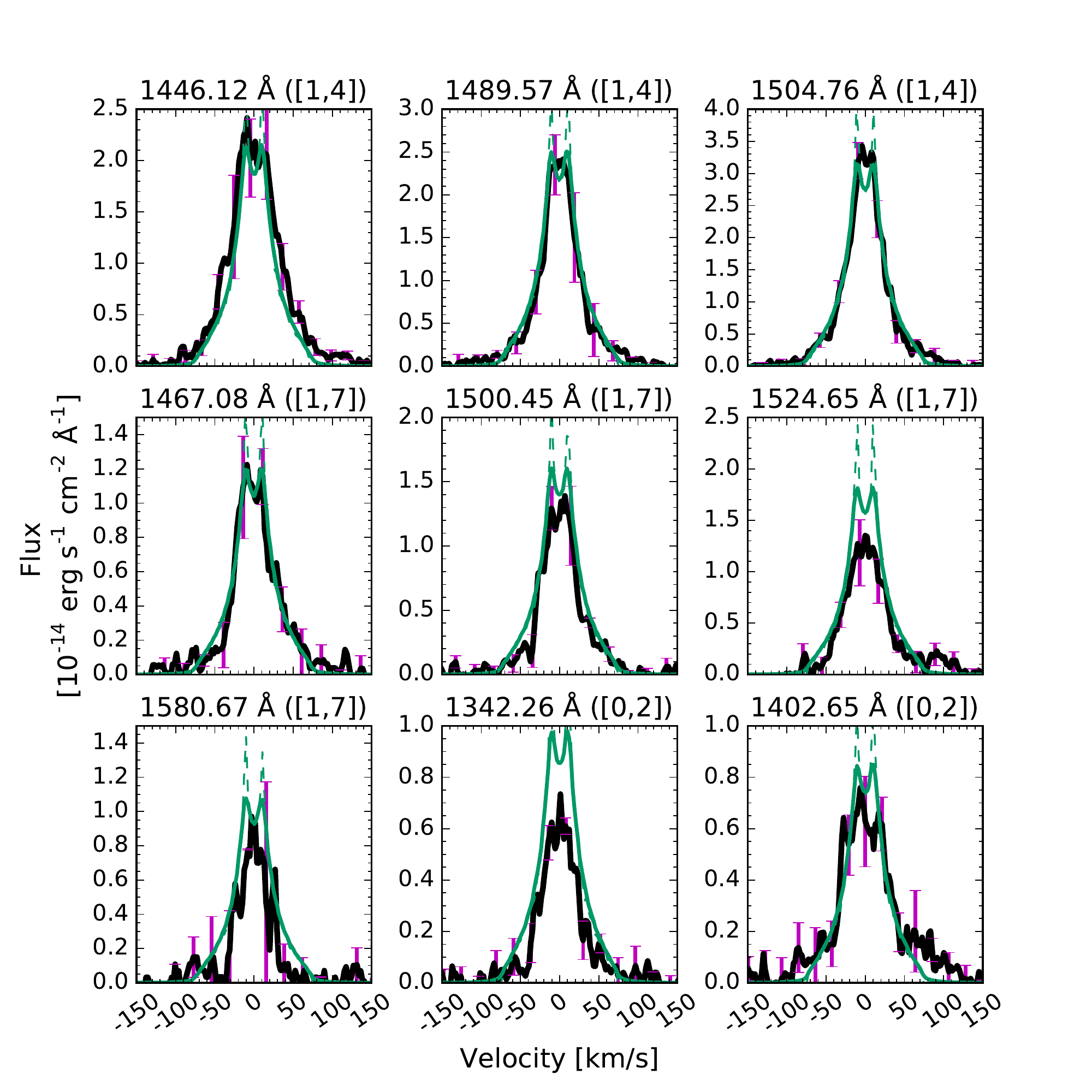}
\caption{Model line profiles (green) of H$_2$ fluorescent emission lines from the [1,4], [1,7], and [0,2] progressions (black), produced with the median values from the posterior distributions of the parameters. All nine lines were fit simultaneously, and the corresponding model parameters inform us about the radial structure of the emitting gas. The dashed green lines show the model line profiles before they were convolved with the \emph{HST}-COS line-spread function.}
\label{H2emissionfits}
\end{figure*}

A Markov chain Monte Carlo (MCMC) routine \citep[emcee;][]{emcee2013} was used to determine the posterior distributions of the model parameters (see Figure \ref{H2emissionfits}). After conducting multiple trials to determine the number of walkers that would allow for convergence while minimizing computation time, we gave the MCMC algorithm a ball of 200 walkers, each at a different set of randomly selected initial conditions, and ran it over 500 steps. The upper and lower limits of the grid space sampled by \citet{Hoadley2015} were used as priors for each parameter, although we restricted the power law index for the temperature distribution, $q$, to values $\geq 0$ to ensure that the temperature either remains constant or decreases with distance from the central star. Best-fit parameters were selected as the median values of the posterior distributions (see Table \ref{H2modelbestparams}). Although the Gaussian fits to the emission lines described in Section 3.1 support the presence of an inner broader component in the line wings, followed by a gap and a second narrower component from larger disk radii, the resolution of our data is not high enough to model this structure under the framework of \citet{Hoadley2015}. However, the detailed modeling approach described here still provides reliable constraints on the radial extent of the hot, fluorescent gas. 

\begin{deluxetable}{cccccc}
\tablecaption{Median Parameters for UV-H$_2$ Emission Line Fits\tablenotemark{a} \label{H2modelbestparams}
}
\tablewidth{0 pt}
\tabletypesize{\scriptsize}
\tablehead{
\colhead{$z/r$} & \colhead{$\gamma$} & \colhead{$T_{1 \, AU}$} & \colhead{$q$} & \colhead{$r_{char}$} & \colhead{$M_{H_2, hot}$} \\
\colhead{$\left[H_p\right]$} & & \colhead{$\left[K \right]$} & & \colhead{$\left[AU \right]$} & \colhead{$\left[M_{\odot}\right]$}
}
\startdata
$2.9 \pm 0.5$ & $1.1^{+0.6}_{-0.5}$ & $1900^{+100}_{-200}$ & $0.40^{+0.05}_{-0.03}$ & $9^{+5}_{-7}$ & $\left(6 \pm 3 \right) \times 10^{-10}$ 
\enddata
\tablenotetext{a}{1-$\sigma$ uncertainties are reported as the values at the 16th and 84th percentiles in each posterior distribution.}
\end{deluxetable}

\begin{figure*}
\centering
\includegraphics[width=0.9\textwidth]
{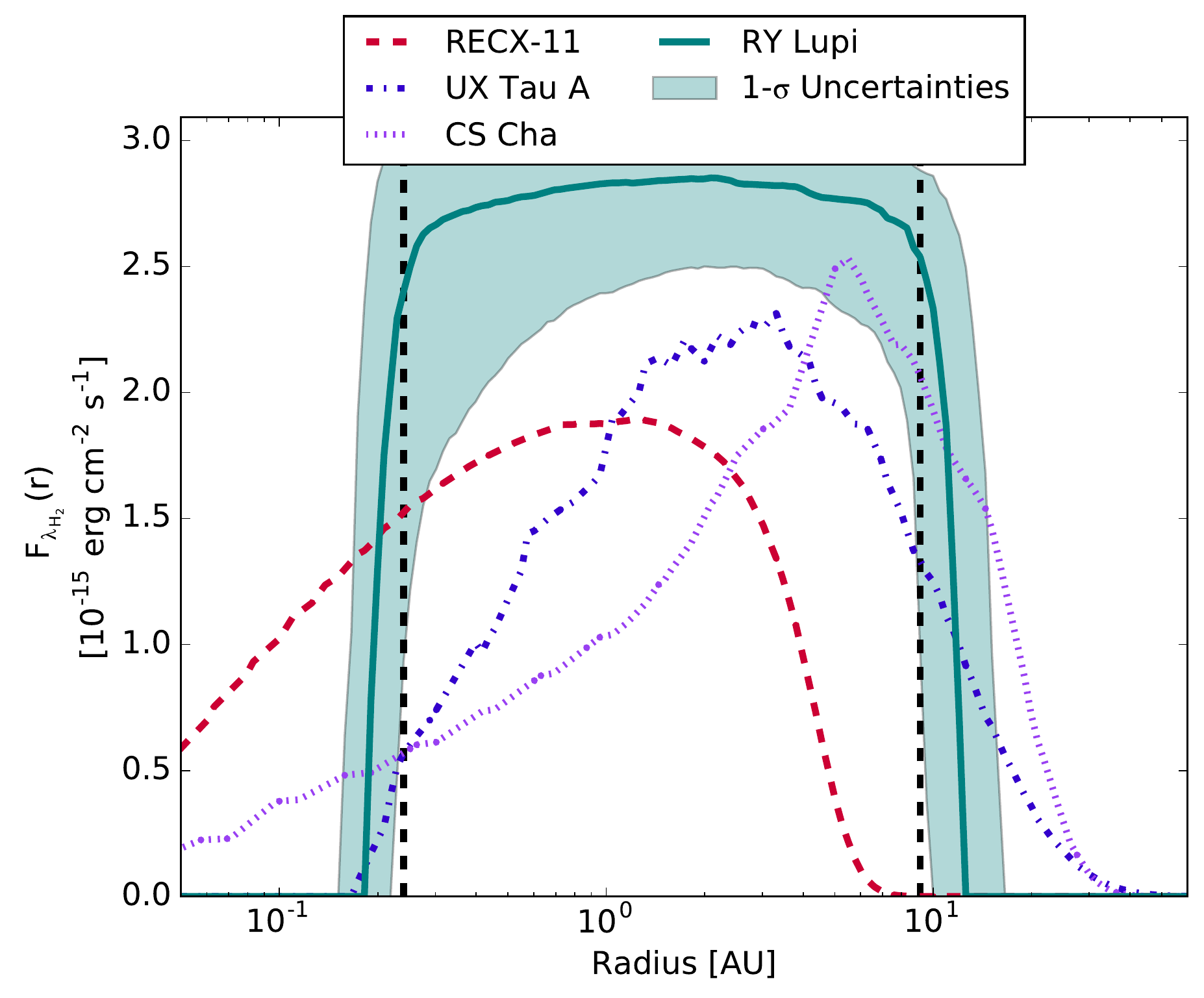}
\caption{Radial distribution of flux from a 2-D radiative transfer model of fluorescent UV-H$_2$ emission lines (turquoise, solid), with 1-$\sigma$ uncertainties (turquoise, shaded). 95\% of the emission from the hot molecular surface layer is enclosed between the black, dashed lines at $r_{in} \sim 0.2$ AU and $r_{out} \sim 9$ AU, making RY Lupi more similar to the sample of systems with dust cavities (e.g. UX Tau A) studied by \citet{Hoadley2015} than the set of full primordial disks (e.g. HN Tau) from the same survey. CS Cha, which has a dust cavity as opposed to a gap like UX Tau A \citep{Espaillat2007_CSCha}, is also shown here for comparison. Note that the flux distributions of UX Tau A, HN Tau, and CS Cha have been scaled down to match the level of the RY Lupi distribution.}
\label{H2fluxdist}
\end{figure*}

The model emission lines correspond to a radial distribution of flux with 95\% of the fluorescent UV-H$_2$ emission from RY Lupi originating between $r_{in} \sim 0.2$ AU and $r_{out} \sim 9$ AU (see Figure \ref{H2fluxdist}). At the $\Delta v \sim 17 \, \text{km s$^{-1}$}$ resolution of \emph{HST}-COS, our observations are sensitive to gas within an average radius of $\sim 17 \, \text{AU}$, as calculated from Equation 2 for a stellar mass of $1.4 \, M_{\odot}$ and an inner disk inclination of $85.6^{\circ}$. This limit implies that the outer radius of $\sim$9 AU from the modeling results is the location of a genuine decline in flux from the hot, fluorescent H$_2$, rather than the detection threshold of the instrument. We concede that the observed H$_2$ emission line profiles do not show the double-peaked shape that would be expected if the outer radial boundary of the hot gas was resolved in the data. As a result, the characteristic radius derived from the modeling results, which describes the location in the disk where the surface density profile changes from a power law distribution to an exponential decline, is rather uncertain. This is reflected in the large 1-$\sigma$ uncertainties on the best-fit value $\left(r_{char} = 9^{+5}_{-7} \, \text{AU} \right)$. Since $r_{char}$ controls the outer extent of the flux distribution, the derived value of $r_{out} \sim 9$ AU may not be robust. However, our observations should be sensitive to gas within $r_{in} = 0.2$ AU, so the inner cutoff in the flux distribution likely represents a physical absence of UV-H$_2$. 

We find that the estimated inner and outer radial bounds of the flux distribution are consistent with the sample of disks studied by \citet{Hoadley2015}, with RY Lupi fitting into the linear trends previously observed for $\mathbf{r_{in}}$ versus mass accretion rate \citep[$\log{ \dot{M}_{acc}} = -8.2$;][]{Alcala2017} and $r_{in}$ versus $r_{out}$. The latter relationship was attributed to an overall outward shift in the distribution of hot H$_2$, with more evolved systems displaying larger values for both $r_{in}$ and $r_{out}$. In the case of RY Lupi, this implies that the clearing of material seen at radii out to $\sim$50 AU \citep{Ansdell2016, vanderMarel2018} may be taking place closer to the star as well. The radial width of the flux distribution from RY Lupi is narrower than what was observed by \citet{Hoadley2015} for most primordial disks, making it similar to the systems from that work with previously detected dust cavities. Our model results also show very little emission inside $r \sim 0.1$ AU, which is roughly consistent with the flux distributions seen by \citet{Hoadley2015} in the gapped disks around TW Hya and LkCa15. Since other young systems have shown inner gas disks extending inside the corotation radius $\left(r_{corot} = 0.05 \, \text{AU for RY Lupi} \right)$, we note that $r_{in}$ does not necessarily trace where the disk has been truncated by a stellar magnetic field \citep{Najita2007}.   

\subsection{4.7 $\mu$m CO Emission Lines}

\begin{figure*}
\centering
\includegraphics[width=0.9\textwidth]
{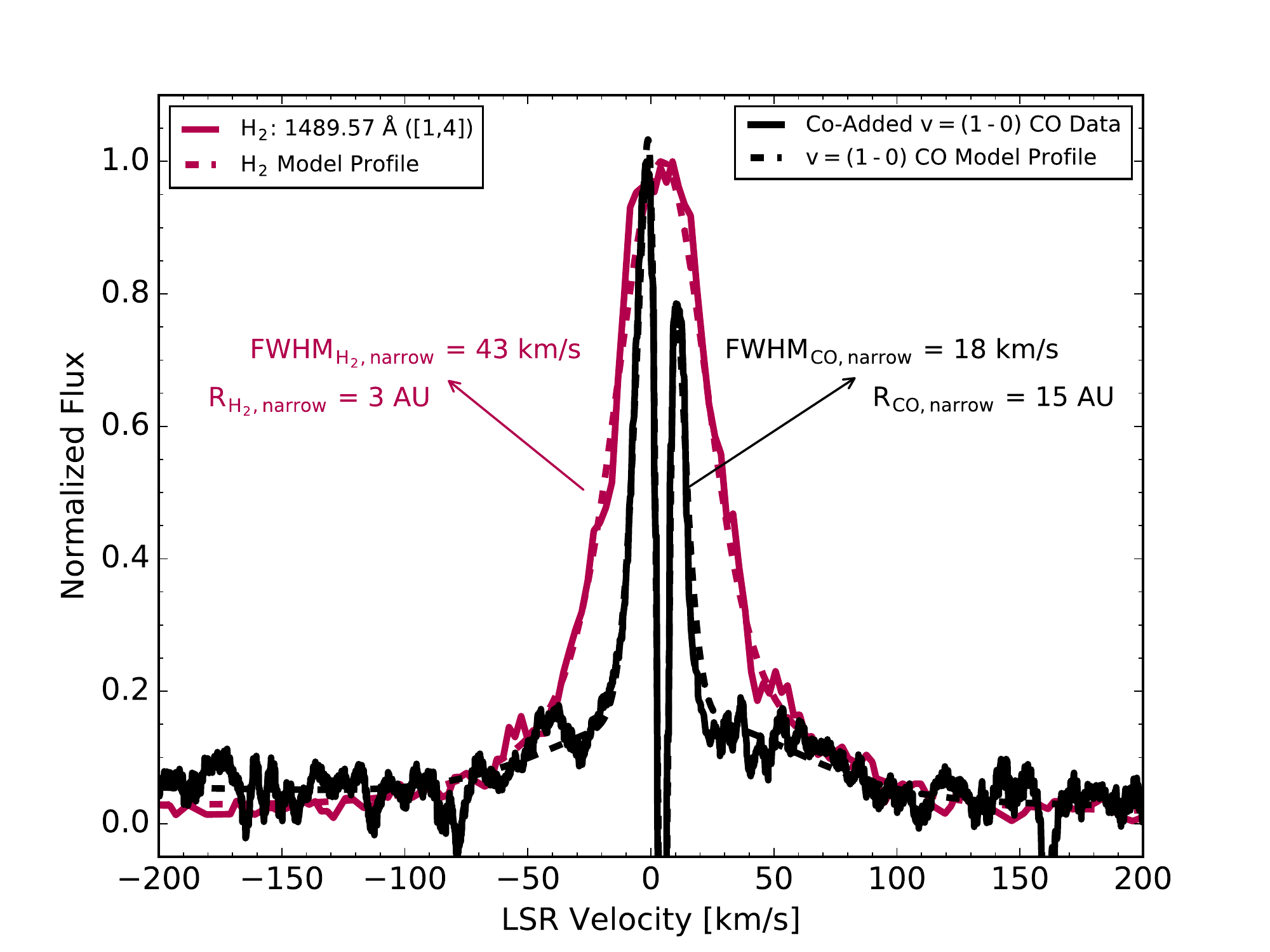}
\caption{The narrow central component of the co-added $\nu = 1-0$ CO profile (black) is narrower than the H$_2$ emission lines (red), showing that the warm CO is located at more distant radii than the hot, fluorescent H$_2$. Note that the parameters for the best-fit Gaussians describing the H$_2$ were obtained after convolving the profiles with the \emph{HST}-COS LSFs.}
\label{H2COcomp}
\end{figure*}

We compare the profiles of the UV-fluorescent H$_2$ to emission from the (1-0) rovibrational transitions of warm CO, which also originate in the system's Keplerian disk (see Figure \ref{H2COcomp}). These infrared lines were observed with VLT-CRIRES $\left( R = 95,000, \Delta v = 3.2 \, \text{km/s} \right)$ in April 2007 and April 2008 \citep{Brown2013}, and the line shapes were classified as emission with broad central absorptions. The blue sides of the lines from the lower rotational states (which are more optically thick) are masked because of telluric absorption, so they were co-added with the features from higher rotational levels to fill in the missing velocity space and increase the S/N (see Table \ref{IRCOemissionlines}). We focus on the CO emission in this section.

The resulting co-added line profile appears to have two velocity components, with a narrow component that is found to be typical in transitional disks \citep{Banzatti2015}. The full profile was modeled as a combination of broad and narrow Gaussian emission components and a central Gaussian absorption. Figure \ref{H2COcomp} compares the co-added CO emission profile to the strongest H$_2$ line from the [1,4] progression. The broad feature has a best-fit FWHM of FWHM$_{broad, CO} = 105 \pm 7$ km s$^{-1}$, which matches the width of the broad H$_2$ emission component measured in Section 3.2. However, the narrow CO component has FWHM$_{narrow, CO} = 18 \pm 1$ km s$^{-1}$, which is significantly smaller than the width of the narrow profile from the two-component fit to the H$_2$ line (FWHM$_{narrow, H_2}$ = 43 $\pm 13$ km s$^{-1}$, after removing the effect of the instrument resolution). Under the model of gas in Keplerian rotation, we use Equation 2 to calculate an average CO radius of $\left \langle r_{narrow, CO} \right \rangle = 15 \pm 2$ AU, which is consistent with the value derived by \citet{Banzatti2017_H2O}. This estimate places the gas roughly 10 AU further out in the disk than the hot H$_2$ that produces the narrow emission line component. We will return to the interpretation of these differences in Section 4.2. 

\begin{deluxetable}{ccc}
\tablecaption{Measured (1-0) CO Infrared Emission Lines \label{IRCOemissionlines}
}
\tablewidth{0 pt}
\tabletypesize{\scriptsize}
\tablehead{
\colhead{Line ID} & \colhead{Wavelength} & \colhead{Oscillator Strength} \\
 & \colhead{(nm)} & \colhead{$\left(10^{-6} \right)$} \\
}
\startdata
(1-0) P(2) & 4682.642826 & 4.64 \\
(1-0) P(3) & 4691.242198 & 4.96 \\
(1-0) P(4) & 4699.949566 & 5.13 \\
(1-0) P(5) & 4708.765629 & 5.24 \\
(1-0) P(6) & 4717.691102 & 5.31 \\
(1-0) P(7) & 4726.726717 & 5.36 \\
(1-0) P(8) & 4735.873214 & 5.39 \\
(1-0) P(11) & 4763.985637 & 5.44 \\
(1-0) P(13) & 4783.295919 & 5.46 \\
(1-0) P(14) & 4793.124102 & 5.46 \\
(1-0) P(17) & 4823.311007 & 5.46 \\
(1-0) P(18) & 4833.610347 & 5.46 \\
(1-0) P(21) & 4865.232183 & 5.45 \\
\enddata
\end{deluxetable}

\subsection{Two Distinct Populations of Absorbing CO}

\begin{figure*}
\centering
\includegraphics[width=0.9\textwidth]
{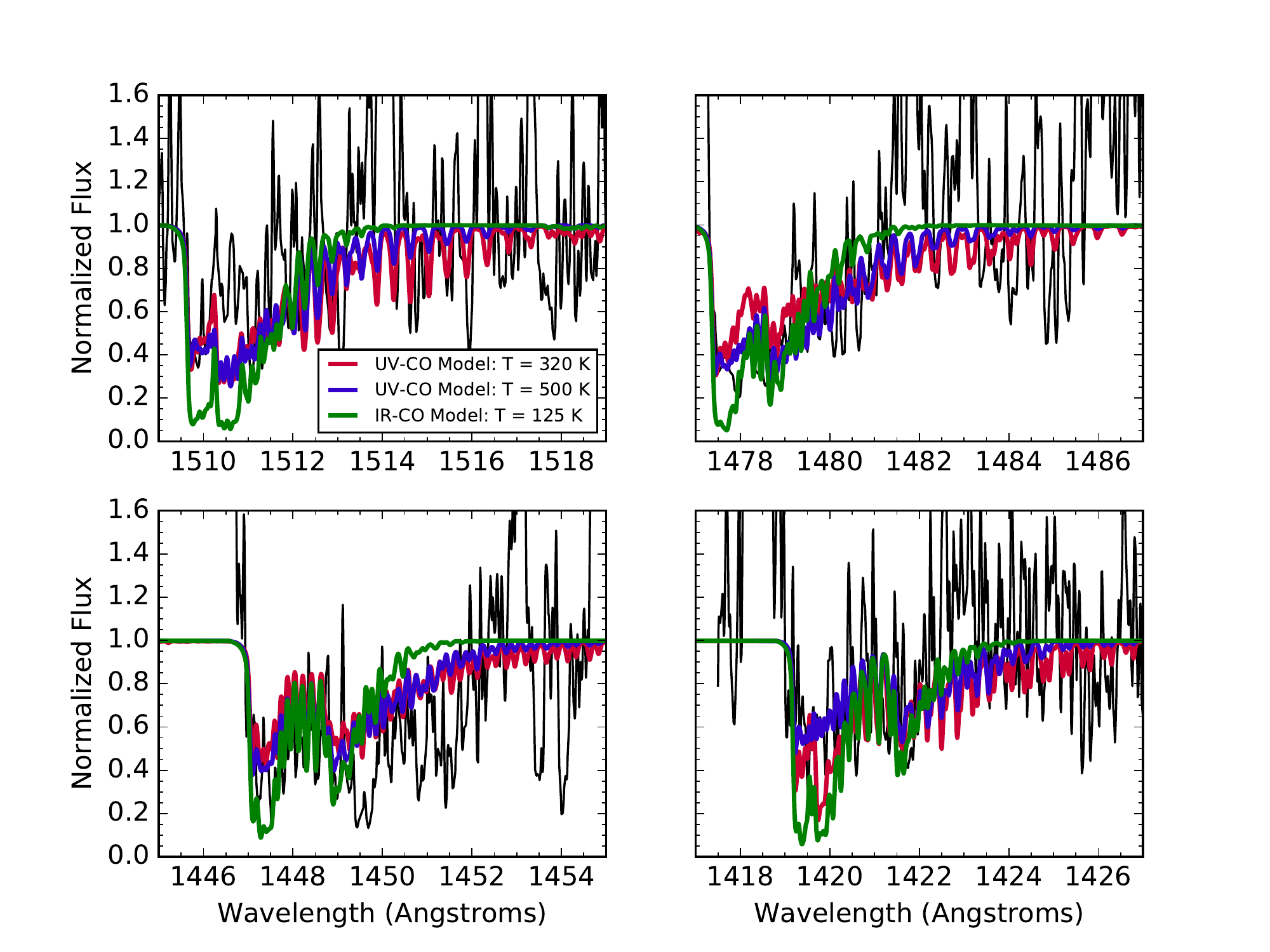}
\caption{Ro-vibrational CO absorptions from the Fourth Positive band system were modeled using the methodology of \citet{McJunkin2013}. The fitting routine identified two different peaks in the posterior distributions (see Table \ref{UV_COresults}), one with $T \sim 500$ K (blue) and one with $T \sim 300$ K (red). Constraints on the $^{12}$CO/$^{13}$CO ratio from the literature \citep{Liszt2007, WoodsWillacy2009} indicate that the model with lower temperature is more physically realistic. We compare the two solutions to the model profile that best represents the IR-CO absorptions (green) and find that it deviates from the UV-CO models for the most prominent lines. However, we note that all of the UV-CO absorption features have low S/N and are therefore not as reliable as the UV-H$_2$ emission lines as tracers of molecular gas in the inner disk.}
\label{UV_CO_bestfitmodels}
\end{figure*}

We present detections of UV-CO absorptions in the Fourth Positive $\left( A^1 \Pi-X^1 \Sigma^+ \right)$ band system and compare them to the 4.7 $\mu$m IR-CO absorptions observed by \citet{Brown2013}. The UV-CO absorption features have been observed in other protoplanetary disks as well \citep{France2011_FUVcontII, McJunkin2013}, although they are not typically present in transitional systems. We have identified UV transitions from $\nu = 0$ to the $\nu' = 1, 2, 3, \text{and } 4$ states in RY Lupi and used the methodology of \cite{McJunkin2013} to generate LTE models of the features. These models allow the Doppler $b$-value, column densities of $^{12}$CO and $^{13}$CO, and gas temperature to float as free parameters. Posterior distributions were derived for each variable using an MCMC routine \citep{emcee2013} consisting of 100 walkers and 500 steps (see Figure \ref{UV_CO_bestfitmodels}). As with the H$_2$ emission lines, the numbers of walkers was chosen to minimize computation time while still allowing the algorithm to converge. 

The MCMC results showed two different solutions for the model parameters describing the warm CO (see Table \ref{UV_COresults}), as expected because of the degeneracy between temperature and column density on the flat part of the curve of growth. However, the value of the $^{12}$CO/$^{13}$CO ratio can be used to constrain which of these solutions is more physically realistic. Within the 1-$\sigma$ uncertainties on the parameter estimates, the model with $T \sim 500$ K, $^{12}\text{CO} \sim 15.5$ (Model 1) has $^{12}\text{CO}/^{13}\text{CO} = 1-3$. The second model, with $T \sim 300$ K and $^{12}\text{CO} \sim 17.2$, has $^{12}\text{CO}/^{13}\text{CO} = 40-250$. Constraints from the literature set bounds of $15 < ^{12}\text{CO}/^{13}\text{CO} < 170$ in the ISM \citep{Liszt2007}, or $25 < ^{12}\text{CO}/^{13}\text{CO} < 77$ in protoplanetary disks \citep{WoodsWillacy2009}. Under these constraints, model 2 represents a more physically realistic environment than model 1. This is also consistent with the disk modeling results of \citet{Miotello2014}, who found that the $^{12}\text{CO}/^{13}\text{CO}$ ratio did not deviate much from their chosen value of 77. However, we note that the low S/N of the data prevents us from placing strong constraints on the physical parameters describing this population of gas. 

\begin{deluxetable}{ccccc}
\tablecaption{Median Parameters for UV-CO and IR-CO Absorption Models \label{UV_COresults}
}
\tablewidth{0 pt}
\tabletypesize{\scriptsize}
\tablehead{
\colhead{Model} & \colhead{T} & \colhead{$\log_{10}$ N($^{12}$CO)} & \colhead{$\log_{10}$ N($^{13}$CO)} & \colhead{b}  \\ 
 & \colhead{(K)} &  &  & \colhead{(km/s)} \\
}
\startdata
UV-CO\tablenotemark{a} & 505$^{+33}_{-20}$ & 15.5 $\pm$ 0.1 & 15.2 $\pm$ 0.1 & 4.9 $\pm$ 1.0 \\
UV-CO\tablenotemark{b} & 320 $\pm$ 20 & 17.2 $\pm$ 0.2 & 15.2 $\pm$ 0.2 & 0.8$^{+1.0}_{-0.8}$ \\
IR-CO & 130 $\pm$ 10 & $16.6 \pm 1$ & - & $2.3 \pm 0.1$ \\
\enddata
\tablenotetext{a}{UV model favored by MCMC results}
\tablenotetext{b}{Degenerate UV model, with parameters estimated from second peak in posterior distribution}
\end{deluxetable}


The sample of targets studied by \cite{McJunkin2013} had best-fit temperatures between 300 and 700 K, a range that encompasses both models for RY Lupi. These estimates are well below the $\sim$1500 K temperature required to produce Ly$\alpha$-pumped fluorescent H$_2$ \citep{Adamkovics2016}, implying that the absorbing UV-CO and emitting UV-H$_2$ are not co-spatial \citep{McJunkin2013, France2014_COH2}. Furthermore, the cooler CO may be as distant as $r \sim 20$ AU \citep{Gorti2008}, depending on the strength of the accretion-dominated stellar UV radiation field. The modeled radial flux distribution from UV-H$_2$ emission is truncated well inside this outer limit (see Figure \ref{H2fluxdist}). 

The models from \citet{McJunkin2013} were adapted to fit nine absorptions from the $\nu = 1-0$ IR band of $^{12}$CO, which were extracted from the centers of the CO emission lines described in Section 3.3. We note that this procedure likely introduced additional uncertainties in the normalized fluxes that are difficult to quantify. To account for these errors in the model fitting procedure, a constant scaling factor was applied to each value. A MCMC run with 500 walkers over 500 steps converged to a best-fit model with $T = 130 \pm 10 \text{ K}$ and $\log_{10}{N\left(^{12}CO \right)} = 16.6 \pm 0.1$. 

\begin{figure*}
\centering
\includegraphics[width=0.9 \textwidth]
{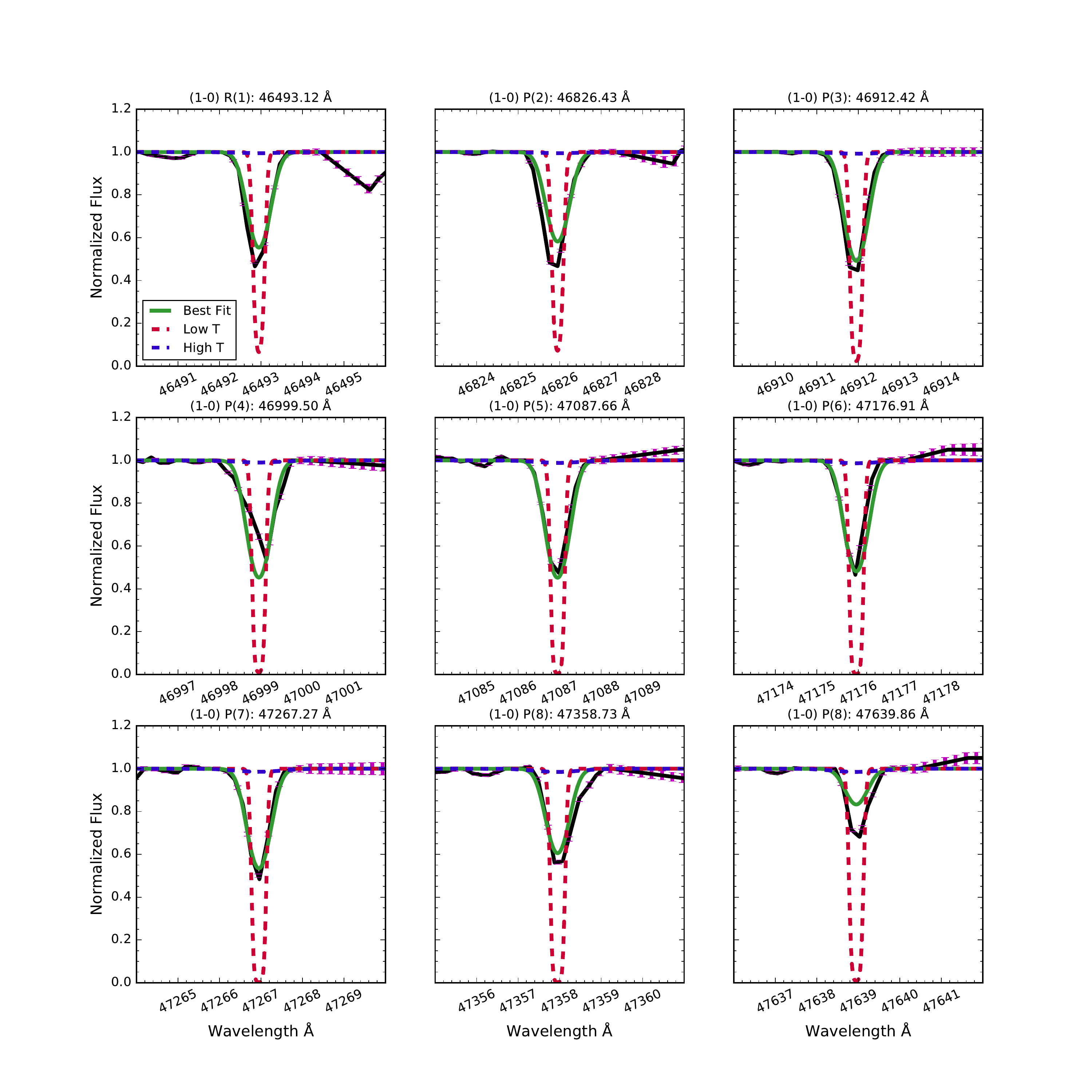}
\caption{Model absorption lines (green) compared to $\nu = \left(1-0 \right)$ IR-CO absorptions near 4.7 $\mu$m (black), along with the low T $\left(T = 300 \text{ K}, N = 17.2 \right)$ and high T $\left(T = 500 \text{ K}, N = 15.5 \right)$ models from the best-fit parameters for the UV Fourth Positive band features. The significant deviations between the IR data and the best-fit UV models further confirm that the two wavelength regimes are probing different populations of gas, although we note again that interpretations of the low S/N UV-CO absorption lines should be taken with caution.}
\label{IRCOabs}
\end{figure*} 

\citet{Brown2013} attributed the IR absorption features to gas in the upper layers of the outer disk, likely at radii more distant than the region probed by the UV data. The median parameters from our IR models are significantly different from both of the solutions we derived for the UV absorptions (see Figure \ref{IRCOabs}), implying that the two populations of absorbing CO are indeed coming from different radii along the line of sight. Figure \ref{IRCOabs} shows that the population of $T \sim 300$ K UV-absorbing CO would have produced deeper IR absorptions than what was observed, implying that the UV-CO is inside the average radius of $\left \langle r_{narrow, CO} \right \rangle \sim 15$ AU derived from the 4.7 $\mu$m emission line peaks. While it is difficult to place any tighter constraints on the physical location of the UV-CO because of the low S/N in the observed absorption lines, the 15 AU radius provides a rough outer limit. 

\subsection{Summary of Results}

\begin{figure*}
\centering
\includegraphics[width=\textwidth]
{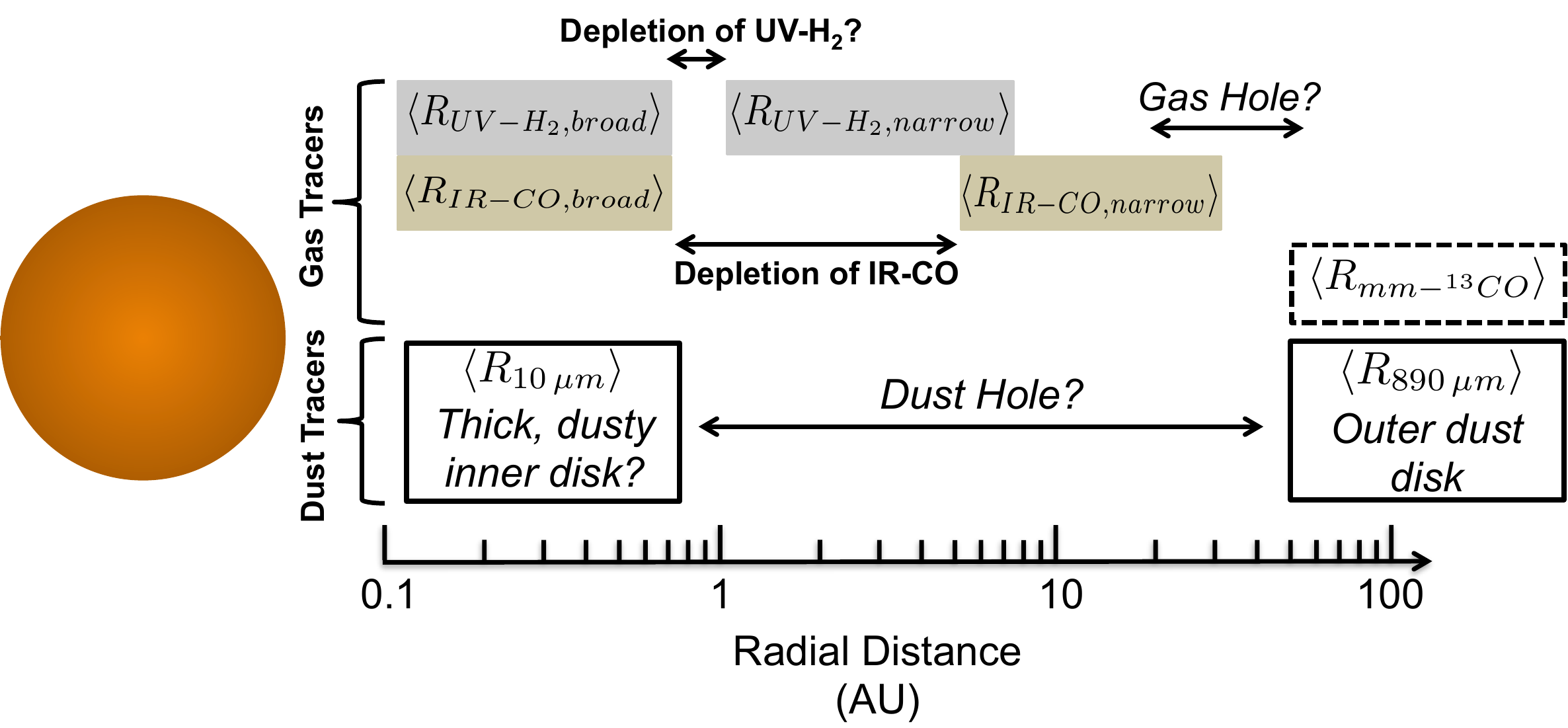}
\caption{A summary of radial structure in the molecular gas disk, showing average distances for UV-H$_2$ $\left( \left \langle r_{broad, H_2} \right \rangle = 0.4 \pm 0.1 \, \text{AU}; \, \left \langle r_{narrow, H_2} \right \rangle = 3 \pm 2 \, \text{AU} \right)$ and IR-CO $\left( \left \langle r_{broad, CO} \right \rangle = 0.4 \pm 0.1 \, \text{AU}; \, \left \langle r_{narrow, CO} \right \rangle = 15 \pm 1 \, \text{AU} \right)$ emission. The mm-$^{13}$CO and 890 $\mu$m dust cavities $\left(r_{cavity} \sim 50 \text{ AU} \right)$ were first observed by \citet{Ansdell2016} and were modeled by \citet{vanderMarel2018}. We also consider the location of dust grains producing the 10 $\mu$m silicate emission feature, which we interpret as evidence for an optically thick inner dust disk. Taken together, these metrics point to the presence of a gas hole in the inner disk, embedded within a larger dust gap extending from $\sim$1-50 AU.}
\label{RYLup_cartoon}
\end{figure*}

Figure \ref{RYLup_cartoon} provides a visual summary of the gas structure within the transitional disk of RY Lupi, as traced by emission from UV-H$_2$ and IR-CO. Both populations of gas are better fit with two-component, rather than single-component, Gaussian profiles. Under the assumption that the two components originate from radially separated regions in a Keplerian disk, we find that the UV-H$_2$ and IR-CO are co-located (although vertically separated) at $\sim$0.4 AU in the inner disk. The second component of UV-H$_2$ emission corresponds to an average gas radius of $\left \langle r_{narrow, H_2} \right \rangle \sim 3$ AU. However, unlike previous studies of these inner disk gas tracers \citep[see e.g.][]{France2012_H2emission}, which have indicated that the UV-H$_2$ and IR-CO probe similar radii, the narrow component of IR-CO emission has a more distant average emitting radius of $\left \langle r_{narrow, CO} \right \rangle \sim 15$ AU. We consider the mechanisms responsible for this discrepancy in Section 4.2, where we also incorporate observational metrics from previous work (10 $\mu$m silicate emission, 890 $\mu$m dust continuum emission, and $^{13}$CO emission) into our discussion of the inner gas disk.

In addition to mapping the radial structure of the gas, we can estimate the relative vertical locations of the emitting UV-H$_2$ and IR-CO. Our assumption that the gas is in LTE requires the kinetic temperature to equal the line temperature \citep[see e.g.][]{Hoadley2015, Schindhelm2012_CO}. This implies that the H$_2$, which must have a temperature of at least $\sim$1500 K for Ly$\alpha$ fluorescence to proceed \citep{Adamkovics2014, Adamkovics2016}, is higher in the disk than the cooler CO \citep[100-1000 K; see e.g.][]{Najita2003}. Although the IR-CO is still close to the surface of the disk, it must sit below the thin layer containing the population of hot UV-H$_2$. 

\section{Discussion}

\subsection{An Inner Disk Warp? Comparison of RY Lupi to AA Tau}

RY Lupi undergoes photometric variations of $\sim$1 mag in the $V$ band over a period of 3.75 days, with an increase in polarization and $B-V$ and $U-B$ colors that become redder when the star is faint \citep{Manset2009}. This behavior was attributed to occultations by a co-rotating dusty warp in the inner disk, much like the geometry previously used to describe AA Tau \citep{Bouvier1999, Menard2003, OSullivan2005, Manset2009}. Recent ALMA observations of AA Tau showed an inclination of $59.1^{\circ} \pm 0.3^{\circ}$ for the outermost dust rings \citep{Loomis2017}, compared to the 75$^{\circ}$ that was previously determined from scattered light measurements \citep{OSullivan2005}. The warp model is no longer able to explain the dimming events in this system if the inner disk is also at this lower inclination, so it is proposed instead that the inner disk is misaligned and closer to edge-on. This effect may also be traced observationally through shadow lanes seen at large radii, which can be modeled to reproduce the opening angle between the inner and outer disks \citep{Marino2015, Min2017, Benisty2017}. A similar geometry may be relevant for RY Lupi, since differing inclination measurements of $i_{outer} = 68^{\circ} \pm 7^{\circ}$ \citep{vanderMarel2018} and $i_{inner} = 85.6^{\circ} \pm 3^{\circ}$ \citep{Manset2009} have been derived from ALMA imaging and scattered light observations, respectively.   

At the time of our observations with \emph{HST}-COS, synthetic photometry from the \emph{HST}-STIS spectrum of RY Lupi showed $V = 12.5$. This magnitude corresponds to a phase where the system was becoming brighter, perhaps as the warp moved out of the line of sight. To examine whether our observations of fluorescent UV-H$_2$ in RY Lupi are sensitive to the inner disk inclination, we adapted the 2-D radiative transfer model described in Section 3.2 and Appendix B to allow the inclination to float as a free parameter. The same MCMC routine described in Section 3.2 was again applied to the UV-H$_2$ data, resulting in a posterior distribution with a median inclination of $i = 72^{\circ} \pm 7^{\circ}$. This value is more consistent with the outer disk inclination, perhaps indicating that the H$_2$ emission is coming from outside the edge-on material probed by the scattered light \citep{Manset2009}. We applied the same inclination-fitting procedure to \emph{HST-}COS spectra of AA Tau as well and found a median inclination of $75^{+4^{\circ}}_{-3^{\circ}}$, which is consistent with the values obtained from other observational signatures of the inner disk \citep{Loomis2017}. This agreement leads us to infer that the best-fit inclination from the RY Lupi UV-H$_2$ is also valid, perhaps providing additional spatial constraints on the strange geometry of this transitional disk. We save further discussion of the warp effect for future work.

\subsection{IR-CO and UV-H$_2$ Radii Indicate Inner Gas Hole}

The discrepancy between the average radii traced by the narrow components of UV-H$_2$ and IR-CO emission is unexpected, since comparisons of these gas tracers in larger samples of circumstellar disks \citep[see e.g.][]{France2012_H2emission} indicate that they probe radially co-located material. To understand why the IR-CO appears to be depleted relative to the UV-H$_2$ around 3 AU, we first consider the mechanisms responsible for producing the observed emission. UV-pumping of CO, analagous to the Ly$\alpha$-pumping of H$_2$, would result in roughly evenly populated states for the $\nu = 1-0$, $\nu = 2-1$ and $\nu = 3-2$ IR-CO bands \citep{Brittain2003}. However, \citet{Banzatti2017_H2O} report a ratio between the second and first vibrational states of $<$ 0.04, indicating that the RY Lupi spectra show no discernible features from the higher rovibrational levels. A similarly low amount of vibrational excitation was also seen in CO spectra of AB Aur \citep{Brittain2003}, a primordial Herbig Ae/Be system which appears to extinguish its UV radiation field at radii much closer to the star than seen in other Herbig systems with similar spectral type. The $\nu = 1-0$ features in AB Aur were attributed to IR rather than UV fluorescence, since the longer-wavelength radiation can penetrate deeper into the disk than the UV continuum. IR photons are much less efficient at exciting the $\nu = 2-1$ and $\nu = 3-2$ bands than the UV continuum, making these transitions fainter than in UV-pumped environments. The lack of strong emission from higher vibrational states is also consistent with collisional excitation in a molecular surface layer \citep{Najita2003}. 

In order for IR fluorescence and collisional excitation to dominate over UV fluorescence, there must be some dust close to the star to attenuate the UV radiation field. Evidence for a residual component of inner shielding dust has recently been identified in a sample of Herbig disks with dust-depleted cavities at larger disk radii (including AB Aur) that still have high NIR excess and very low CO vibrational ratios despite their strong UV radiation \citep{Banzatti2017_ABAur}. These disks also show evidence for inner dust belts that may be misaligned or warped compared to the outer disk \citep[e.g. ][]{Benisty2017, Min2017}. Such inner warped disk structure may explain the observed emission in RY Lupi, too, as discussed in Section 4.1. This is further supported by the strength of the 10 $\mu$m silicate emission feature in RY Lupi, which indicates that there is still dusty material present in the inner regions of the disk despite the observed dust cavity radius of $\sim$50 AU \citep{Ansdell2016, vanderMarel2018}. Given the luminosity of RY Lupi \citep[$L_{\ast} = 2.6 \, L_{\odot}$;][]{Bouvier1990}, we use the relationship
\begin{equation}
\log R = -0.45 + 0.56 \log \left(L_{\ast} / L_{\odot} \right) 
\end{equation}
derived by \citet{KS2007} to calculate a silicate emission radius of 0.6 AU, which potentially marks the rim of a large dust hole between the inner and outer disks. The broad CO component may be emitted from a UV-shielded region just beyond the inner dust belt, as suggested for AB Aur by \citet{Brittain2003}. 

We note that our estimate of a silicate emission radius at 0.6 AU is rather uncertain. The data used by \citet{KS2007} to derive the relationship between stellar luminosity and silicate emission radius show a large amount of scatter, which those authors attribute to variations in disk geometry that arise when considering systems that may be in different stages of dust evolution. It is possible that the 10 $\mu$m silicate emission instead originates in small, warm grains \citep{Espaillat2007_UXTauLkCa15} distributed somewhere within the observed mm-wave cavity \citep{vanderMarel2018}. This optically thin region could either extend all the way in to the sublimation radius \citep[e.g. CS Cha;][]{Espaillat2007_CSCha} or separate the outer disk from an optically thick inner disk \citep[e.g. LkCa 15, UX Tau A, ROX 44;][]{Espaillat2010_pretransitional}. Although detailed modeling of the near-to-mid-IR SED of RY Lupi may be required to definitively distinguish between these two scenarios, we can use the observations of UV-H$_2$, UV-CO, and IR-CO presented in this work to establish a preferred geometry.

A 10 $\mu$m silicate emission feature originating from an optically thin distribution of dust with no optically thick wall to shield it is the preferred model for the disk around CS Cha, which \citet{Espaillat2007_CSCha} placed at a more evolved stage of evolution than the systems with optically thick inner disks like LkCa 15 and UX Tau A. All three of these objects were included in the UV-H$_2$ survey of \citet{Hoadley2015} and show very different distributions of hot, Ly$\alpha$-pumped gas. Those authors find that CS Cha has UV-H$_2$ emission coming from more distant radii than any other disk in their sample, with 95\% of its flux distribution contained within an outer radius of $\sim$22 AU. By contrast, the flux distributions of LkCa 15 and UX Tau A only extend out to 6 and 12 AU, respectively, consistent with RY Lupi's outer radius of 9$^{+5}_{-7}$ AU. This suggests that UV photons do not penetrate as far into the circumstellar environment as they do in CS Cha, perhaps because the radiation field is partially truncated by an optically thick inner disk.

Thermal emission from an inner disk has been identified as the cause of veiled near-IR photospheric features from LkCa 15 and UX Tau A \citep{Espaillat2010_pretransitional}. The excess continuum flux fills in the stellar absorption lines, causing them to appear weaker than expected based on the spectral type of the star. This is commonly seen in systems with full, primordial disks \citep{Espaillat2010_pretransitional}. Although we have not analyzed the full near-IR spectrum of RY Lupi in this work, we note that the excess flux observed in the system's SED led to its classification as a primordial disk \citep{KS2006}. Since the shape of the UV-H$_2$ flux distribution in RY Lupi is also more similar to the less evolved systems, like LkCa 15 and UX Tau A, we favor the description of the 10 $\mu$m silicate feature as optically thin emission arising in a region between optically thick inner and outer disks. This is further supported by the detection of UV-CO absorptions in RY Lupi (presented in Section 3.4), which have not yet been detected in disks with dust cavities \citep{McJunkin2013}. We note that the gap and cavity models are both consistent with the ALMA data presented in \citet{vanderMarel2018}, which do not have sufficient resolution to distinguish between the two scenarios.

Under the assumption of a dusty inner disk truncated somewhere outside of $\left< r_{broad, CO} \right> = \left< r_{broad, H_2} \right> = 0.4$ AU, we may expect to see a drop in the surface density of gas at this distance as well. However, molecules are still expected to survive within the dust cavity if the column density of gas is large enough for self-shielding or if the rim of the dust disk is high enough to block some of the radiation field \citep{Bruderer2013_cavitygas, Bruderer2014_IRS48}. Since we detect emission from Ly$\alpha$-pumped UV-H$_2$ at $\left< r_{narrow, H_2} \right> \sim 3$ AU, the obscuration must not entirely shield the hot gas from UV photons. The H$_2$ is able to self-shield and will continue to produce UV emission lines until the gas layer is too cool for Ly$\alpha$ pumping to proceed \citep{Adamkovics2014, Adamkovics2016}. This same H$_2$ should shield the CO from photodissociation as well, so the observed depletion of IR-CO cannot be attributed to the lack of dust-shielding alone. 

A build-up of gas is expected to occur at the inner edge of the dust cavity \citep{Bruderer2013_cavitygas, vanderMarel2013, Bruderer2014_IRS48}, which would allow the CO molecules to once again produce $\nu = 1-0$ emission. However, we observe IR-CO emission at $\left< r_{narrow, CO} \right> \sim 15$ AU, which is well inside the observed dust cavity radius of $\sim$50 AU \citep{Ansdell2016, vanderMarel2018}. It is possible that the narrow component of IR-CO emission is produced from a build-up of gas just inside an as-yet-unresolved dust ring. Alternatively, the $\left< r_{narrow, CO} \right> \sim 15$ AU radius could trace the location where the CO has accumulated to a large enough column density for self-shielding \citep{Bruderer2014_IRS48}. 

We interpret $\left< r_{narrow, CO} \right>$ as a rough estimate of the inner radius of a gas hole in the disk around RY Lupi. The outer edge of the hole is located at the $\sim$50 AU radius traced by the dust continuum and mm observations of $^{13}$CO, in agreement with the models of \citet{vanderMarel2018} at the resolution of the ALMA data. This type of gap is more consistent with clearing by a protoplanet(s) located near $\left(\left< r_{narrow, CO} \right> \sim 15 \text{ AU} \right)$ than by grain growth or photoevaporation alone \citep{Bruderer2014_IRS48}. Furthermore, photoevaporation, which dissipates the disk from the inside out \citep[see e.g.][]{vanderMarel2013}, is unlikely to leave a residual optically thick inner disk \citep{Espaillat2008_LkCa15proof}, implying again that planet formation may be a more plausible mechanism for dust clearing.

\subsection{Symmetric H$_2$ Line Profiles Show No Signs of a Disk Wind}

\begin{figure*}
\centering
\includegraphics[width=0.9 \textwidth]
{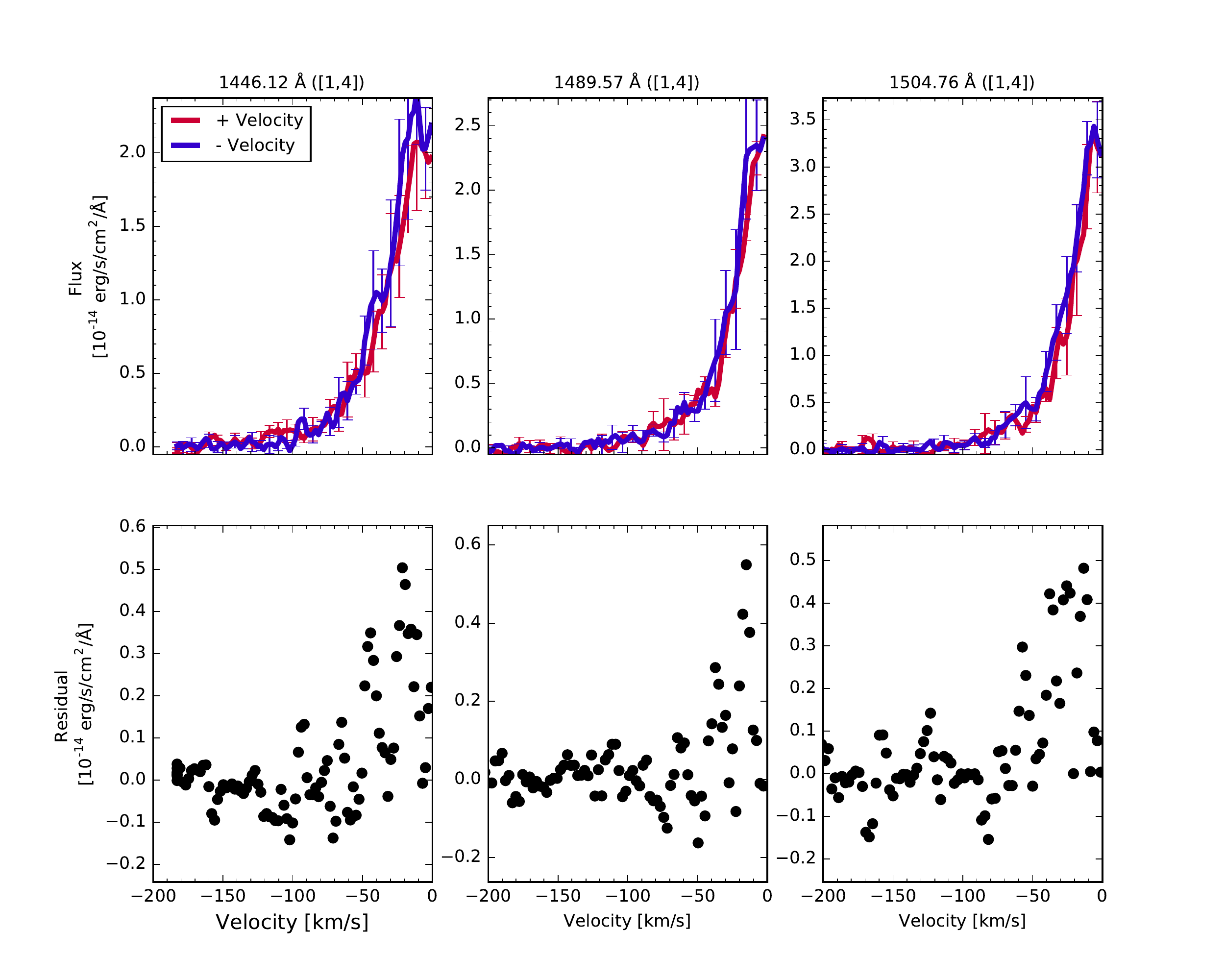}
\caption{Mirrored profiles of H$_2$ emission lines from the [1, 4] progression. An Anderson-Darling test was conducted for each feature to determine whether differences between the red and blue sides of the line profiles are statistically significant $\left(p < 0.05 \right)$. At the resolution of our data, there are no meaningful asymmetries.}
\label{mirroredlines}
\end{figure*}

Additional H$_2$ emission from a protostellar outflow or a warm disk wind could produce asymmetries between the red and blue sides of the line profiles \citep{Herczeg2006, France2012_H2emission, Hoadley2015}. We investigated this effect in RY Lupi by mirroring the red halves of the lines and overplotting them on the blue sides, as shown in Figure \ref{mirroredlines} \citep{Pascucci2011_mirror}. A two-sample Anderson-Darling (A-D)\footnote{The A-D test was chosen in place of a Kolmogorov-Smirnov (K-S) test, because the K-S test statistic is not sensitive to deviations at the tails of the distributions (see references in \citet{Feigelson2012}). Since the excess emission could be present in the peak or wings of the line profile, a K-S test may overlook these deviations. The A-D test was designed to maintain better sensitivity across the entire probability distribution, making it a good replacement for the standard K-S test in this case.} test statistic was computed for each feature to examine whether the red and blue sides of the line profile represent similar probability distributions. We found that only the (1-7) P(8) 1524.65 \AA \, and (0-6) P(2) 1460.17 \AA \, H$_2$ emission lines had A-D test statistics that were significant at the $p < 0.05$ level (see Table \ref{mirroringresults}), and the (1-7) P(8) 1524.65 \AA \, feature is contaminated by the (0-7) P(3) 1525.15 \AA \, H$_2$ line. Three other profiles had $p < 0.10$, but otherwise no convincing asymmetries were detected. However, we note the trend in the residuals that indicates stronger blue emission near the peaks of the line profiles. To check whether this difference was significant, we calculated A-D test statistics from data at $v < 15$ km/s only, finding that the resulting test statistics were insignificant at the resolution of our data. We conclude that there are no detectable signatures of excess emission due to a molecular outflow or hot disk wind in this spectrum. 

\begin{deluxetable}{ccc}
\tablecaption{Anderson-Darling Test Statistics for Asymmetries in H$_2$ Emission Line Profiles \label{mirroringresults}
}
\tablewidth{0 pt}
\tablehead{
\colhead{Progression} & \colhead{Wavelength (\AA)} & \colhead{Statistic}\tablenotemark{a} \\
}
\startdata
[1,4] & 1446.12 & -0.127 \\
 & 1489.57 & -0.240 \\
 & 1504.76 & -0.201 \\
\hline
[1,7] & 1467.08 & 1.79 \\
 & 1500.45 & 2.01 \\
 & 1524.65 & 4.62 \\
 & 1580.67 & -0.122 \\
\hline
[0,1] & 1338.56 & -0.0184 \\
 & 1398.95 & 1.36 \\
 & 1460.17 & 9.92 \\
\hline
[0,2] & 1342.26 & 0.402 \\
 & 1402.65 & 1.31 \\
\enddata
\tablenotetext{a}{Critical values of the test statistic are 1.226, 1.961, and 3.752, corresponding to $p=0.1$, $p=0.05$, and $p=0.01$. Very few of our measured statistics are larger than the critical values, implying that there is no statistically significant difference between the red and blue halves of the line profiles.}
\end{deluxetable}         

\section{Summary and Conclusions}

We have presented \emph{HST}-COS and \emph{HST}-STIS observations of the gas disk around the young star RY Lupi, which show UV-H$_2$ fluorescent emission and UV-CO $A-X$ band absorptions. A radial distribution of flux from the UV-H$_2$ was estimated by modeling the Ly$\alpha$-pumped emission lines, and the results show a hot gas distribution with 95\% of its flux coming from between $r_{in} = 0.2$ AU and $r_{out} = 9$ AU. A closer examination of the shapes of these emission features shows that they are better fit by a two-component line profile, produced by two radially separated rings as opposed to a smooth distribution of gas. We note that our current framework for modeling the radial distribution of flux from the UV-H$_2$ emission lines is unable to reproduce the two-component distribution at the resolution of our data. However, the bounds of the flux distribution $\left(r_{in}, \, r_{out} \right)$ from our model of a smooth inner disk still provide useful constraints on the spatial extent of the emitting gas. 

For the first time, the UV-H$_2$ data were interpreted in conjunction with 4.7 $\mu$m IR-CO emission lines, which are also well-represented by a two-component model. The two populations of gas traced by the broad line component are co-located in the innermost regions of the disk, but the warm CO appears to be depleted relative to the hot H$_2$ beyond $\sim$0.4 AU. When we consider this result in the context of the 890 $\mu$m dust continuum, 10 $\mu$m silicate emission, and $^{13}$CO mm emission, we find evidence of a gas hole in the inner disk, embedded within a larger dust gap that extends from $\sim$1-50 AU. This makes RY Lupi similar to the systems LkCa 15, UX Tau A, and GM Aur, which have optically thick inner and outer disks that are separated by large gaps. However, the IR-CO \citep{Najita2003, Salyk2011, Brown2013} and UV-H$_2$ emission from these three objects appear to trace radially co-located material in their gas disks, in keeping with the general trend observed by \citet{France2012_H2emission}. RY Lupi's deviation from this behavior implies that it may be in a slightly more or less evolved phase than these other systems, making it an important environment for testing models of inner disk evolution.  

As an additional probe of the complex structure within the disk around RY Lupi, we used the UV-H$_2$ fluorescence models to determine the inclination of the emitting region. The result is more consistent with observations of the outer disk than with the value obtained from polarization measurements of scattered light in the inner disk, implying that the bulk of the UV-H$_2$ emission originates outside the highly inclined inner disk. We also place an upper limit on the radius of UV-CO observed in absorption, which  must come from inside $\left \langle r_{narrow, CO} \right \rangle \sim 15 \, \text{AU}$. Taken together, the UV and IR datasets have allowed us to piece together the radial and vertical structure within the dust cavity around RY Lupi, as traced by different gas parcels. The panchromatic approach used to study RY Lupi in this work will be extended to three other systems in the Lupus complex with different disk morphologies (MY Lup, Sz 68, and TYC 7851), two of which appear to have full primordial disks and one that has a more evacuated 890 $\mu$m dust cavity. After characterizing the warm and hot molecular gas in the inner regions of these disks, we will be left with a sample of co-evolving objects that will help us construct a more empirically motivated picture of disk evolution. 

We would like to thank the referee for their comments, which greatly enhanced the discussion presented here. We are also grateful to M. Tazzari, J.S. Pineda, R. Loomis, and K. Flaherty for helpful discussions about RY Lupi. This work was supported by \emph{HST}-GO program 14469. NA is supported by NASA Earth and Space Science Fellowship grant 80NSSC17K0531. CFM acknowledges support from an ESA Research Fellowship, and CFM and AM acknowledge support from ESO Fellowships. Astrochemistry in Leiden is supported by the Netherlands Research School for Astronomy (NOVA), by a Royal Netherlands Academy of Arts and Sciences (KNAW) professor prize, and by the European Union A-ERC grant 291141 CHEMPLAN. AB acknowledges support by \emph{HST}-GO program 14703. JMA acknowledge financial support from the project PRIN-INAF 2016 The Cradle of Life - GENESIS-SKA (General Conditions in Early Planetary Systems for the rise of life with SKA). This research made use of Astropy, a community-developed core Python package for Astronomy \citep{astropy2013, astropy2018}.

\begin{appendices}

\section{\\ Error Handling in Low S/N Data} \label{App:AppendixA}
\setcounter{table}{0}
\renewcommand{\thetable}{A\arabic{table}}

The \emph{HST}-COS data reduction pipeline is optimized for spectra from sources with high S/N ratios, where it is assumed that the number of counts detected follows a Poisson distribution. However, the Poisson uncertainties are incorrectly treated as symmetric Gaussian errors, which results in anomalously low continuum errors for sources like RY Lupi with low S/N (fluxes roughly $<10^{-15}$ erg s$^{-1}$ cm$^{-2}$ \AA$^{-1}$). These values are not necessarily reflective of the true uncertainty in the data and will give points in the continuum a misleadingly large weight when fitting models. 

We use a Markov chain Monte Carlo (MCMC) method to determine the posterior distributions of the model parameters (Foreman-Mackey et al. 2013), which easily allows us to adopt a weighting scheme that approximates the Poisson behavior of the data. We assume that each data point $\left(x_i \right)$ is drawn from a normal distribution with mean $\mu_i$ and uncertainty $\sigma_i$. The log-likelihood function for the low S/N data is then
\begin{align}
\mathcal{L} \left( \vec{x} | \vec{\mu}, \vec{\sigma} \right) &= -0.5 \times \sum_{i=1}^n \left[ \frac{\left( x_i - \mu_i \right)^2}{s_i^2} + \ln \left(s_i^2 \right) \right] \\
s_i^2 &= \sigma_i^2 + \left(f^2 \times \mu_i^2 \right), \\
\end{align}
where an additional model parameter $\left(f \right)$ is incorporated to account for the fractional underestimate in our uncertainties from the \emph{HST-}COS pipeline. A second extra model parameter was used to determine a flux threshold, which sets the level below which the uncertainties in the data are underestimated. All fluxes above this threshold were added to the log-likelihood function without correction, since the error bars associated with high-signal data points are more representative of the measurement uncertainty.  

\section{\\ 2-D Radiative Transfer Models of Ly$\alpha$-Pumped Fluorescent H$_2$} \label{App:AppendixB}
\setcounter{table}{0}
\renewcommand{\thetable}{B\arabic{table}}

The 2-D radiative transfer models from \citet{Hoadley2015} assume that the fluorescent UV-H$_2$ emission originates in a thin surface layer of gas in Keplerian rotation, with ground state populations in local thermodynamic equilibrium (LTE). The physical structure of the emitting gas is described by a radial temperature distribution (Section 3.2, Equation 5)
\begin{equation*}
T \left( r \right) = T_{1 \, AU} \left( \frac{r}{1 \, \text{AU}} \right)^{-q},
\end{equation*}
pressure scale height 
\begin{equation}
H_p \left( r \right) = \sqrt{ \frac{k T \left( r \right)}{\mu m_H} \frac{r^3}{G M_{\ast}}},
\end{equation}
and surface density distribution (Section 3.2, Equation 6)
\begin{equation*}
\Sigma \left(r \right) = \Sigma_c \left( \frac{r}{r_c} \right)^{- \gamma} \exp \left[ - \left( \frac{r}{r_c} \right)^{2-\gamma} \right]
\end{equation*}
that are assumed to be independent of height in the thin gas layer. The surface density normalization constant 
\begin{equation}
\Sigma_c = \frac{M_{H_2} \left( 2- \gamma \right)}{X_{H_2} \left(2 \pi r_c^2 \right)}
\end{equation}
is dependent on the power-law index $\gamma$, the mass of the H$_2$ in the disk $\left(M_{H_2} \right)$, the fraction of $M_{H_2}$ that contributes to the emission $\left(X_{H_2} \right)$, and a characteristic radius $\left( r_c \right)$ beyond which the gas distribution is dominated by exponential decay. The first five model parameters that we estimate from the fitting routine $\left( T_{1 \, AU}, \, q, \, r_c, \, \gamma, \, \text{and} \, M_{H_2} \right)$ are all used to constrain this radial structure of the hot, fluorescent layer. 

The sixth model parameter $\left( z / r \right)$ is used to calculate the mass density
\begin{equation}
\rho \left(r, z \right) = \frac{ \Sigma \left( r \right)}{\sqrt{2 \pi} H_p} \exp \left[-0.5 \left( \frac{z}{H_p \left(r \right)} \right)^2 \right],
\end{equation} 
which then allows us to determine the number density of H$_2$ in ground state $\left[ v, J \right]$ from the Boltzmann equation:
\begin{equation}
n_{\left[v, J \right]} \left(r, z \right) = \frac{ \rho \left(r, z \right) X_{H_2}}{\mu m_H} \times \frac{g_{\left[v, J \right]}}{Z_{\left[v, J \right]} \left(T \right)} \times \exp \left( - \frac{E_{\left[v, J \right]}}{k T \left( r \right)} \right).
\end{equation}
This number density sets the optical depth $\tau_{\lambda} \left(r, z \right)$ of the gas, which controls the amount of flux
\begin{equation}
F_{\lambda_{H_2}} = \eta S_{\lambda} \left(r, z \right) \times B_{mn} \sum \limits^{\tau'_{\lambda}} \left(1 - e^{- \tau'_{\lambda} \left(r, z \right)} \right)
\end{equation}  
from each H$_2$ emission line, with branching ratio $B_{mn}$. The source function $\left( S_{\lambda} \left(r, z \right) \right)$ in Equation 14 is represented by the Ly$\alpha$ flux, while $\eta$ describes what fraction of the Ly$\alpha$ is intercepted by molecular gas. We assume that each H$_2$ line is isotropically emitted from each individual parcel in the hot layer, such that the flux reaching the observer is given as
\begin{equation}
F_{\lambda_{H_2}} = \eta F_{\ast, Ly\alpha} \left( \frac{R_{\ast}^2}{r^2} \right) \left( \frac{ \left(d \cos i_{disk} \right)^2}{s \left(r, z \right)^2} \right) \times B_{mn} \sum \limits^{\tau'_{\lambda}} \left(1 - e^{- \tau'_{\lambda} \left(r, z \right)} \right)
\end{equation}
We sum these fluxes over the entire disk to produce final emission line profiles that can be fit to the data. 

\end{appendices}

\bibliographystyle{apj}
\bibliography{RYLup_bibliography}

\end{document}